\documentclass[11pt,tightenlines,floats,groupedaddress,sort&compress,merge,superscriptaddress,a4,twocolumn]{revtex4}

\usepackage[latin9]{inputenc}
\usepackage[letterpaper]{geometry}
\usepackage{verbatim}
\usepackage{amsmath}
\usepackage{amssymb}
\usepackage{graphicx}
\usepackage[unicode=true,
 bookmarks=false,
 breaklinks=false,pdfborder={0 0 1},backref=false,colorlinks=false]
 {hyperref}
\usepackage{breakurl}

\makeatletter
\@ifundefined{textcolor}{}
{%
 \definecolor{BLACK}{gray}{0}
 \definecolor{WHITE}{gray}{1}
 \definecolor{RED}{rgb}{1,0,0}
 \definecolor{GREEN}{rgb}{0,1,0}
 \definecolor{BLUE}{rgb}{0,0,1}
 \definecolor{CYAN}{cmyk}{1,0,0,0}
 \definecolor{MAGENTA}{cmyk}{0,1,0,0}
 \definecolor{YELLOW}{cmyk}{0,0,1,0}
}

\usepackage{bbding}
\usepackage{xcolor}
\usepackage{amsthm}\usepackage{amsfonts}\usepackage{enumitem}\usepackage{dsfont}
\usepackage{yfonts}\usepackage{calligra}
\usepackage{graphicx}
\usepackage{pxfonts}
\usepackage{upgreek}%\usepackage{lmodern}
\usepackage{geometry}
\providecommand{\U}[1]{\protect\rule{.1in}{.1in}}

\makeatother

\begin{document}

\title{Towards practical device-independent quantum key distribution with
spontaneous parametric downconversion sources, on-off photodetectors
and entanglement swapping}

\author{Kaushik P. Seshadreesan}

\affiliation{National Institute of Information and Communications Technology,
Koganei, Tokyo 184-8795, Japan}

\affiliation{Hearne Institute for Theoretical Physics and Department of Physics
and Astronomy, Louisiana State University, Baton Rouge, Louisiana
70803, USA}

\affiliation{Max-Planck-Institut f\"{u}r die Physik des Lichts,
G\"{u}nther-Scharowsky-Straße 1, Bau 24, 91058 Erlangen, Germany }

\author{Masahiro Takeoka}

\affiliation{National Institute of Information and Communications Technology,
Koganei, Tokyo 184-8795, Japan}

\author{Masahide Sasaki}

\affiliation{National Institute of Information and Communications Technology,
Koganei, Tokyo 184-8795, Japan}

\date{\today}
\begin{abstract}
Device-independent quantum key distribution (DIQKD) guarantees unconditional
security of secret key without making assumptions about the internal
workings of the devices used. It does so using the loophole-free violation
of a Bell's inequality. The primary challenge in realizing DIQKD in
practice is the detection loophole problem that is inherent to photonic
tests of Bell\textquoteright s inequalities over lossy channels. We
revisit the proposal of Curty and Moroder {[}Phys. Rev. A 84, 010304(R)
(2011){]} to use a linear optics-based entanglement-swapping relay
(ESR) to counter this problem. We consider realistic models for the
entanglement sources and photodetectors; more precisely, (a) polarization-entangled
states based on pulsed spontaneous parametric downconversion (SPDC)
sources with infinitely higher order multi-photon components and multimode
spectral structure, and (b) on-off photodetectors with non-unit efficiencies
and non-zero dark count probabilities. We show that the ESR-based
scheme is robust against the above imperfections and enables positive
key rates at distances much larger than what is possible otherwise.
\end{abstract}
\maketitle

\section{Introduction}

Quantum cryptography \cite{SBCDLP09,GRTZ02} uses the laws of quantum
mechanics to establish unconditional security of data transmission---meaning
that the encrypted data can be secure against an eavesdropper of unbounded
abilities. The BB84 \cite{BB84} and a host of other protocols proposed
since \cite{E91,B92,GG02,SARG04,LMC05} guarantee such unconditional
security in quantum key distribution (QKD) when the physical components
used are well characterized and trustworthy. However, such ideal conditions
cannot be met perfectly in the real world. The implementation of the
physical devices may have imperfections more or less, i.e., side channels.
Also, the components may have been manufactured by a malicious party,
introducing backdoors into them. Real world quantum crypto-systems
are hence amenable to a plethora of possible attacks through side
channels and backdoors. This has stimulated great interest in a model
for cryptography that establishes security independently of the internal
workings of the physical devices used and is thus inherently immune
to side-channel attacks and backdoors, provided that the given devices
are operated in secure locations by the legitimate sender (Alice)
and receiver (Bob) \cite{MY98}. Such a ``device independent'' (DI)
model for QKD has been carefully studied and its security proven under
fairly general conditions (cf. \cite{VV14,PABGMS09,ABGMPS07} and
references therein).

\begin{figure}
\includegraphics[scale=0.52]{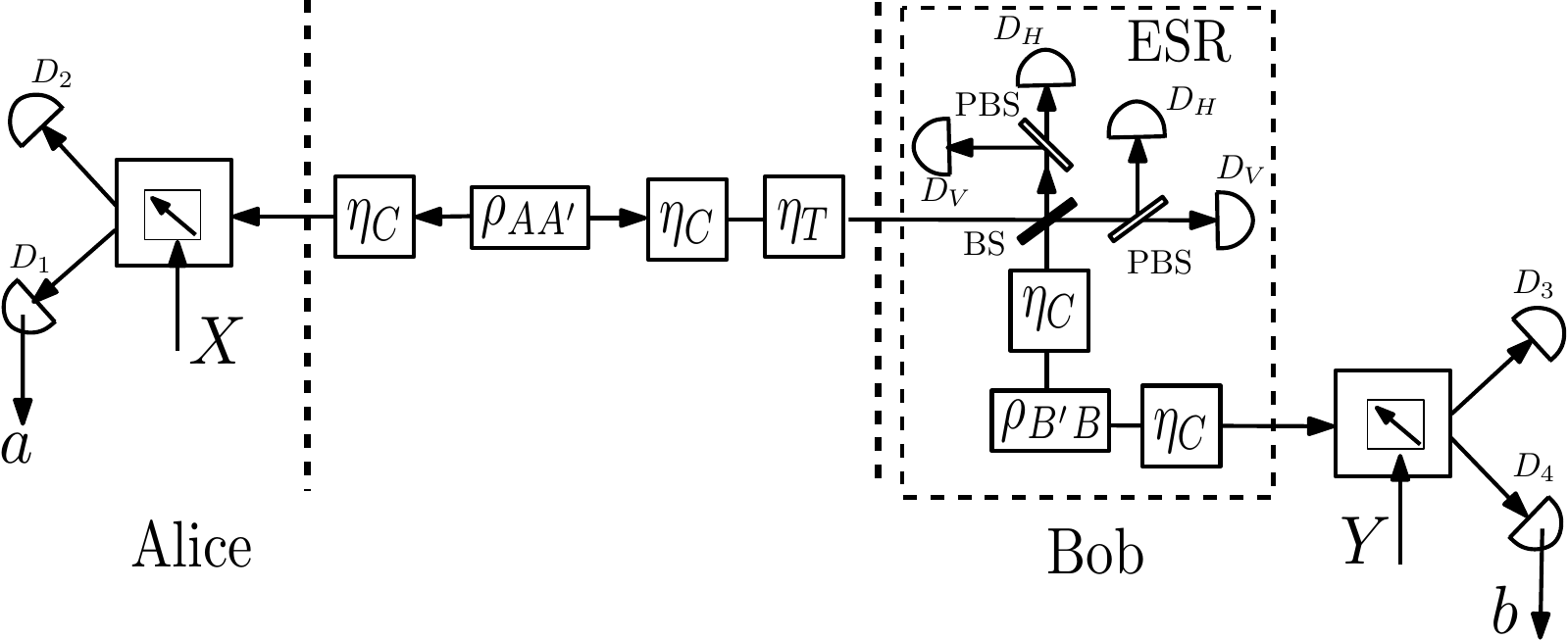}

\caption{Setup for DIQKD with a conventional entanglement-swapping relay (ESR)
node based on linear optics. A source $\rho_{AA'}$ distributes polarization
entanglement to receivers Alice and Bob. The distributed states are
subject to losses in fiber coupling (both in the channel to Alice
as well as to Bob) and transmission (in the channel to Bob, who is
situated far from the source), the respective efficiencies being $\eta_{T}$
and $\eta_{C}$. Bob employs a ESR node, which consists of another
similar entanglement source $\rho_{BB'}$, beam splitters (BS: 50:50
beam splitter, PBS: Polarizing beam splitter) and heralding detectors
$D_{H}$ and $D_{V}$ corresponding to horizontal and vertical polarizations.
Upon each successful entanglement swapping event, Alice and Bob perform
polarization measurement with polarizer settings $X$, $Y$, respectively,
and the outcomes are denoted as $a,\:b\in\left\{ +1,-1\right\} $,
respectively. The detectors are assumed to be imperfect, on-off photodetectors,
with non-unit efficiencies and non-zero dark-count probabilities.\label{fig:schematic}}
\end{figure}

Unconditional security in DIQKD is typically guaranteed by means of the loophole-free
violation of a Bell's inequality, where both the locality and the
detection loopholes are closed simultaneously \cite{BCPSW14}. The
first-ever loophole-free Bell test has been performed with electron
spins in nitrogen vacancy (NV) centers in diamonds \cite{Henson15}.
The first-ever all-optical loophole-free Bell tests \cite{GVW15,SMC15}
have also been realized recently. Yet, photonic Bell tests over long-distance
communication channels are bound to suffer from the detection loophole
problem due to transmission and fiber-coupling losses. Nevertheless,
there have been proposals to mitigate transmission losses using non-deterministic
strategies. In particular, inspired by Ralph and Lund's idea for a
non-deterministic photon amplifier \cite{RL09}, Gisin et al. \cite{GPS10}
proposed a heralded qubit amplifier that utilizes quantum teleportation
to boost the amplitude of the maximally entangled component of a lossy
entangled state. The qubit amplifier was demonstrated experimentally
by Kocsis et al. \cite{Lvov13,KXRP13}. It is, however, technically
far from feasible for application in DIQKD. Curty and Moroder \cite{CM11}
investigated a conventional entanglement-swapping relay (ESR) node
based on linear optics (Fig. \ref{fig:schematic}). Rather than amplifying
the maximally entangled component in the lossy state, the relay node
simply ensures that the state heralded upon successful entanglement
swapping sufficiently violates the Clauser-Horne-Shimony-Holt (CHSH)
inequality \cite{CHSH69} in a loophole-free test. The authors showed
that the relay node enables higher key rates than what is possible
with the teleportation-based qubit amplifier when photon number resolving
detectors (PNRD) are used and the product of coupling and detector
efficiencies is higher than 95\%. High efficiency entanglement swapping
has been successfully demonstrated in numerous optical experiments
\cite{JTTSS15,HBGSSZ07,PBWZ98}. More recently, in an alternative approach, DIQKD
based on local Bell tests has been considered and its security investigated \cite{LPTRG13}.

In this work, we revisit the scheme of Curty and Moroder based on entanglement swapping \cite{CM11} with realistic models for the entanglement sources and detectors.
We consider sources of polarization-entanglement, which are based
on a pair of pulsed SPDCs with infinitely higher order multi-photon
components and multimode spectral structure. Pulsed sources are preferred
over continuous-wave sources in many practical applications, because
they generate temporally localized signals which are more suitable
for photon counters and coincidence count measurements. On the other
hand, these signals are generated in spectrally multiple modes, which
makes it more difficult to match the mode of interest between the
devices used, especially when dispersive components are involved in
the experimental setup. Thus, in order to improve the visibility of
correlation measurements, careful multimode analysis is necessary.
This is our motivation to include the multimode spectral structure
of the sources in our model. We model our detectors as on-off photodetectors---detectors
that merely distinguish the event of presence of photons from absence,
and include losses and dark counts. The detector efficiencies are
assumed to be flat over all the spectral modes. We show that the relay
node enables positive key rates at distances larger than what is possible
without the relay node for sufficiently large coupling and detector
efficiencies, small dark count probabilities in the detectors and
small spectral spread in the sources. Our analyses are non-perturbative
and exact. They involve the use of tools from Gaussian quantum information
that are based on characteristic functions.

The paper is organized as follows. In Section \ref{sec:ESR}, we recall
the basics of DIQKD and outline the ESR-assisted scheme for DIQKD.
In Section \ref{sec:Modeling-Imperfections}, we describe our realistic
model for the sources of polarization entanglement in the scheme,
which includes their higher order multi-photon components and multimode
spectral structure. In Section \ref{sec:Nonlocality-ESR}, we present
our results. Section \ref{sec:Discussion-and-Conclusion} captures
our main conclusions.%
\begin{comment}
In Section \ref{sec:Characteristic-Function-Approach}, we present
a brief review of the characteristic function approach from Gaussian
quantum information. Here, we also discuss our model for the photodetectors,
and models for the imperfections in the channel and detectors, such
as transmission loss, free-space to fiber coupling loss, detector
loss, and detector dark counts.
\end{comment}
{}

\section{DIQKD using an Entanglement-Swapping Relay \label{sec:ESR}}

\subsection{Basic principle of DIQKD}

First of all, we recall the basic principle of DIQKD between two parties
Alice and Bob \cite{AMP06}. A typical protocol for DIQKD involves:
(a) a ``blackbox'' source that transmits shares of an entangled
quantum state to Alice and Bob through lossy communication channels,
and (b) a blackbox measurement apparatus at each of Alice and Bob.
The apparatus at Alice has three possible measurement settings $X_{i}\in\{X_{0},X_{1},X_{2}\}$,
while the one at Bob has two possible settings, namely $Y_{j}\in\{Y_{1},Y_{2}\}$.
All the measurement observables are taken to have binary outcomes,
i.e., $a_{i},\:b_{j}\in\left\{ +1,-1\right\} $. For example, in an
optical protocol for DIQKD based on polarization entanglement, the
different measurement settings would correspond to different polarizer
settings, and the outcomes to the clicking of one of two detectors
placed in orthogonal polarization modes. The only assumption involved
is that Alice and Bob are in secure locations such that no classical
information either about the choice of measurement settings or the
observed outcomes leaks out without their permission.

Alice and Bob perform repeated measurements under the setting $\left\{ X_{0},Y_{1}\right\} $
to generate the raw key. The qubit error rate (QBER) associated with
the raw key is defined as $P\left(a\neq b\left|X_{0}=Y_{1}\right.\right)$.
Over a subset of uses of the communication channel, Alice and Bob
use the measurement settings $\left\{ X_{1},X_{2}\right\} $ and $\left\{ Y_{1},Y_{2}\right\} $
to test the CHSH functional
\begin{equation}
\mathcal{CHSH}=\left\langle a_{1}b_{1}\right\rangle +\left\langle a_{1}b_{2}\right\rangle +\left\langle a_{2}b_{1}\right\rangle -\left\langle a_{2}b_{2}\right\rangle ,\label{eq:chsh}
\end{equation}
where $\left\langle a_{i}b_{j}\right\rangle =P\left(a=b\left|X_{i}Y_{j}\right.\right)-P\left(a\neq b\left|X_{i}Y_{j}\right.\right)$.
A value of $\mathcal{{CHSH}}>2$ indicates the presence of nonlocal
correlations in the state and is used to bound Eve's knowledge about
the key. We denote the maximal possible value of $\mathcal{CHSH}$
for a given state with the corresponding sets of optimal measurement
observables $\left\{ X_{1},X_{2}\right\} $ and $\left\{ Y_{1},Y_{2}\right\} $
by $S$. $S$ can at best take the value $2\sqrt{{2}}$, known as
the Cirelson bound \cite{Cirel80}, and is achieved by the maximally
entangled state. The key rate is a function of $S$ and the QBER.
A conservative lower bound on the rate of generating key that is secure
against the so-called collective eavesdropping attacks (i.e., where
the attack is independent and identical during each use of the communication
channel) is given by the Devetak-Winter formula \cite{DW05}:
\begin{equation}
K\geq1-h\left(Q\right)-\chi\left(S\right),
\end{equation}
where $K$ is the number of secret bits that can be generated per
channel use%
\begin{comment}
$P_{s}$ is the probability of successful heralding at the relay node
marked by the clicking of any one pair of detectors $D_{6}$ and $D_{7}$
or $D_{5}$ and $D_{8}$ and the respective other pair not clicking
\end{comment}
, $S$ is the maximal violation, $Q$ is QBER, %
\begin{comment}
given by
\begin{align}
Q & =\min\bigg\{ P\left(a\neq b\left|\theta_{X}=\theta_{Y}\right.\right),\\
 & P\left(a\neq b\left|\theta_{X}=\theta_{Y}'\right.\right)\bigg\},
\end{align}
where $\theta_{Y}$ and $\theta_{Y}'$ are the polarizer settings
corresponding to $Y=0$ and $Y=1$, respectively (at Bob), 
\end{comment}
\begin{equation}
\chi\left(S\right)=h\left(\frac{{1+\sqrt{{\left(S/2\right)^{2}-1}}}}{2}\right),
\end{equation}
and $h\left(x\right)$ is the binary entropy given by $h\left(x\right)=-x\log x-\left(1-x\right)\log\left(1-x\right)$.
It can be shown that $S>2$ is a necessary condition to realize a
positive key rate. %
\begin{comment}
Although $S>2$ is a necessary condition to guarantee unconditionally
secure secret key, it is not sufficient to identify if key can be
distilled at a meaningful rate.
\end{comment}

A crucial requirement on the Bell test for DIQKD is that it is performed
in a loophole-free manner. We recall a simple strategy that has been
used to perform a loophole-free test of the CHSH inequality for the
realistic scenario under consideration \cite{CM11}. Let us say that
the clicking of Detector $D_{1}$ at Alice corresponds to outcome
$a=+1$ and $D_{2}$ to $a=-1$ and similarly, the clicking of $D_{3}$
at Bob to $b=+1$ and $D_{4}$ to $b=-1$. When neither or both detectors
at Alice ($D_{1}$, $D_{2}$) click, or likewise at Bob ($D_{3}$,
$D_{4}$), the outcome is obviously inconclusive at the respective
party. The strategy is to deterministically assign a conclusive outcome
upon such detection events. For example, when neither or both detectors
at Alice click, the outcome can be assigned the value $a=-1$, and
at Bob $b=-1$.

\subsection{Entanglement-Swapping Relay-assisted DIQKD}

Suppose that the source of entanglement is located near Alice, and
Bob is situated at a distance from both Alice and the source. The
losses in the communication channel to Alice are thus attributed to
fiber coupling and detector inefficiencies. On the other hand, the
channel to Bob in addition suffers from transmission losses. We denote
the fiber-coupling efficiency, the detector efficiency, and the transmission
efficiency by $\eta_{C}$, $\eta_{\textrm{det}}$ ( or $\eta_{\operatorname{hdet}}$
in the case of heralding detectors) and $\eta_{T}$, respectively.
We refer to the product $\eta_{C}\eta_{\textrm{det}}$ as detection
efficiency $\eta_{D}$ (or $\eta_{HD}$ in the case of the heralding
modes), while on the other hand, by ``overall'' detection efficiency,
we mean the product $\eta_{D}\eta_{T}$. The overall detection efficiencies
at Alice and Bob are thus given by $\eta_{D}$ and $\eta_{D}\eta_{T}$,
respectively. Recent results by Caprara-Vivoli et al. \cite{VSBLC15}
have shown that a loophole-free test of the CHSH inequality in (\ref{eq:chsh})
based on the deterministic strategy described above requires an overall
detection efficiency, which is at least 2/3 to exhibit a value of
$S>2$. Assuming ideal fiber coupling and detectors at both parties
(i.e., $\eta_{D}=1$), this corresponds to a distance of $8.8$km
for the optical fiber communication channel to Bob ($\alpha=0.2$dB/km
attenuation). Since $S>2$ is a necessary condition for a positive
key rate, the scope for DIQKD thus appears to be severely limited
at first look. However, as mentioned before \cite{GPS10,CM11}, it
is possible to mitigate the effects of transmission losses on distillable
key rate using probabilistic strategies, thereby extending the possible
distances for DIQKD.

Consider the ESR-assisted scheme shown in Fig \ref{fig:schematic},
as considered in Ref. \cite{CM11}. To first approximation, the state
generated by the source $\rho_{AA'}$ is a maximally polarization
entangled photon pair, with a photon directed towards each of Alice
and Bob. Alice performs polarization measurement on her share of the
state. Bob, on his part, employs an ESR node to the state received
through the lossy channel. That is, he mixes the incoming state on
a 50:50 beamsplitter with one share of another polarization entangled
state $\rho_{B'B}$, which is similar to $\rho_{AA'}$ and performs
polarization measurement on the output modes. When entanglement swapping
succeeds, (i.e., when either of the pair of heralding detectors $D_{6}$
and $D_{7}$ or $D_{5}$ and $D_{8}$ placed in the output modes click
and the respective other pair doesn't) Bob performs a polarization
measurement on the other share of the entangled state $\rho_{B'B}$.
The parties then apply the deterministic strategy of assigning conclusive
values to inconclusive outcomes mentioned above to perform DIQKD based
on the (loophole-free) CHSH test. Naturally, the key rate in the ESR-assisted
scheme now includes a factor corresponding to the probability of success
of the relay node. Although this success probability drops exponentially
with growing distance, the distances over which positive key rate
can be achieved with ideal fiber coupling and detectors still improves
by an order of magnitude compared to the original scheme.

\section{Modeling our Entanglement Sources\label{sec:Modeling-Imperfections}}

We now describe our realistic model for the ESR-assisted DIQKD scheme
discussed above that includes imperfections. All detectors are modeled
as on-off photodetectors, i.e., they simply distinguish between vacuum
and not vacuum. The model takes into account dark count probability
and non-unit efficiency (see Appendix for more details). The sources
of polarization entanglement are modeled using realistic SPDCs. A
detailed account of the same is described below.

\subsection{Polarization entanglement based on a pair of SPDC sources\label{sub:Polarization-entanglement}}

Polarization entangled photon pairs form a natural choice for entangled
qubits in photonic implementations of QKD. For example, one could
consider generating a photon-pair state of the form 
\begin{equation}
\propto\left(\left|H_{A},V_{B}\right\rangle +\left|V_{A},H_{B}\right\rangle \right),
\end{equation}
where the polarization of the photons that Alice and Bob receive are
oppositely correlated. One way to achieve such an entangled state
in practice is by using a pair of weakly pumped SPDC sources, as described
below.

\begin{figure}
\includegraphics{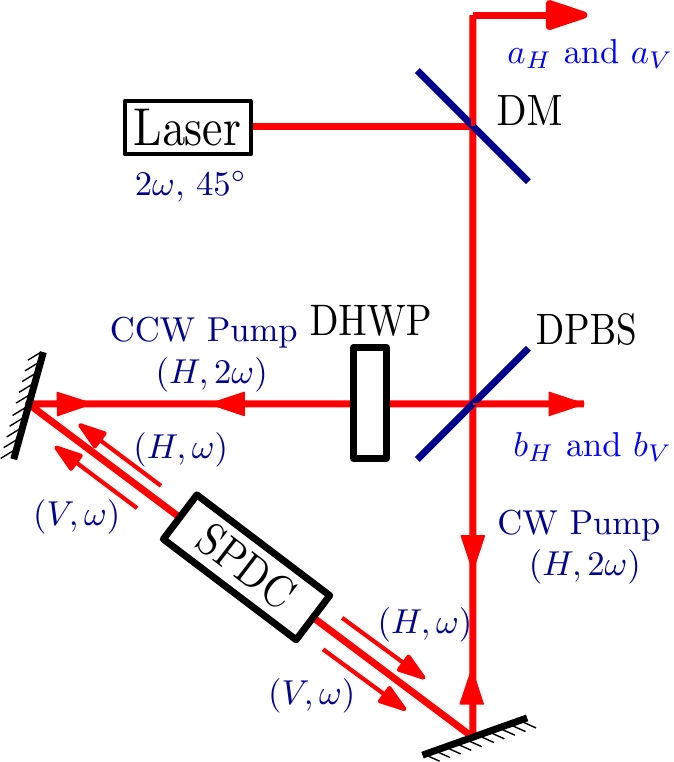}

(a)

\bigskip{}

\includegraphics{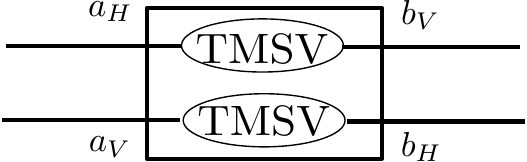}

(b)

\caption{\label{fig:source}(a) A Sagnac loop source for generating polarization
entanglement based on Type-II SPDC, meaning it produces downconverted
light in two orthogonal polarization modes. The SPDC is pumped simultaneously
by a clockwise (CW) and a counter-clockwise (CCW) pump. DM stands
for a dichroic mirror, which reflects light of frequency $2\omega$,
while being transparent to light of frequency $\omega$. DPBS stands
for a dichroic polarizing beamsplitter, which splits light of both
frequencies $\omega$ and $2\omega$ into its $H$ and $V$ polarization
components. DHWP stands for a dichroic halfwave plate at an angle
$45^{\circ}$, so that it flips the polarization in the $H,V$ basis
as $H\rightarrow V$ and $V\rightarrow H$. The geometry of the source
makes sure that the pump is perfectly recycled, while the down-converted
light is output through the topmost and the rightmost modes. (b) A
schematic representation of the source. The CW pump generates squeezing
in the modes $a_{H}$ and $b_{V}$, while the CCW pump generates squeezing
in the other two modes.}
 
\end{figure}

Consider a Sagnac loop architecture with a single nonlinear crystal,
with the crystal being pumped simultaneously from both clockwise (CW)
and counter-clockwise (CCW) directions as shown in Figure \ref{fig:source}
(a). The crystal is assumed to enable Type-II SPDC, meaning it produces
downconverted light in two orthogonal polarization modes. The resulting
state can be described as follows. Let us denote the input modes to
the Sagnac loop as modes $\hat{{a}}$ and $\hat{b}$. Then, the state
at the output of the two SPDC processes (CW and CCW) can be approximated
as $\left|\Psi\right\rangle $
\begin{align}
=\left(c_{0}\left|00\right\rangle _{a_{H}a_{V}}+c_{1}\left|11\right\rangle _{a_{H}a_{V}}\right)\left(c_{0}\left|00\right\rangle _{b_{H}b_{V}}+c_{1}\left|11\right\rangle _{b_{H}b_{V}}\right)\nonumber \\
=c_{0}^{2}\left|0000\right\rangle _{a_{H}a_{V}b_{H}b_{V}}+c_{1}^{2}\left|1111\right\rangle _{a_{H}a_{V}b_{H}b_{V}}\nonumber \\
+c_{0}c_{1}\left(\left|0011\right\rangle _{a_{H}a_{V}b_{H}b_{V}}+\left|1100\right\rangle _{a_{H}a_{V}b_{H}b_{V}}\right),
\end{align}
where we assume $\left|c_{0}\right|^{2}+\left|c_{1}\right|^{2}\approx1$
and $\left|c_{1}\right|\ll$1 and $a_{H}$, $a_{V}$ denote the horizontal
and vertical polarization modes in the spatial mode $\hat{{a}}$,
for example. This state, when propagated through the polarizing beamsplitter,
results in 
\begin{align}
\left|\Psi\right\rangle \xrightarrow{\operatorname{PBS}} & c_{0}^{2}\left|0000\right\rangle _{a_{H}a_{V}b_{H}b_{V}}+c_{1}^{2}\left|1111\right\rangle _{a_{H}a_{V}b_{H}b_{V}}\nonumber \\
+ & c_{0}c_{1}\left(\left|0110\right\rangle _{a_{H}a_{V}b_{H}b_{V}}+\left|1001\right\rangle _{a_{H}a_{V}b_{H}b_{V}}\right).
\end{align}
When post-selected on its two-photon component, this state gives the
desired polarization entangled photon pair \cite{JSWF14}. The underlying
essence behind the generation of entanglement here is the lack of
information as to which of the two pumps resulted in the generation
of the downconverted photon pairs. Since the underlying source of
the photon pair in the considered scheme is a pair of SPDC processes,
the source can be exactly modeled as a tensor product of two-mode
squeezed vacuums
\begin{align}
\exp\left(\xi\hat{a}_{H}^{\dagger}\hat{b}_{V}^{\dagger}-\xi^{*}\hat{a}_{H}\hat{b}_{V}\right)\exp\left(\xi\hat{a}_{V}^{\dagger}\hat{b}_{H}^{\dagger}-\xi^{*}\hat{a}_{V}\hat{b}_{H}\right)\nonumber \\
\left|0\right\rangle _{a_{H}}\otimes\left|0\right\rangle _{a_{V}}\otimes\left|0\right\rangle _{b_{H}}\otimes\left|0\right\rangle _{b_{V}}\label{eq:source-tmsv}
\end{align}
where, e.g., $\hat{a}_{H}$ and $\hat{a}_{H}^{\dagger}$ are the annihilation
and creation operators of the mode $a_{H}$. Figure \ref{fig:source}
(b) depicts this representation of the state produced by the source.

We note that alternatively one could simply use a single downconversion
source and also achieve polarization entanglement of the same merit.
We choose the model based on two downconversions, because it is a
convenient option in experiments.

\begin{figure*}
\includegraphics[clip]{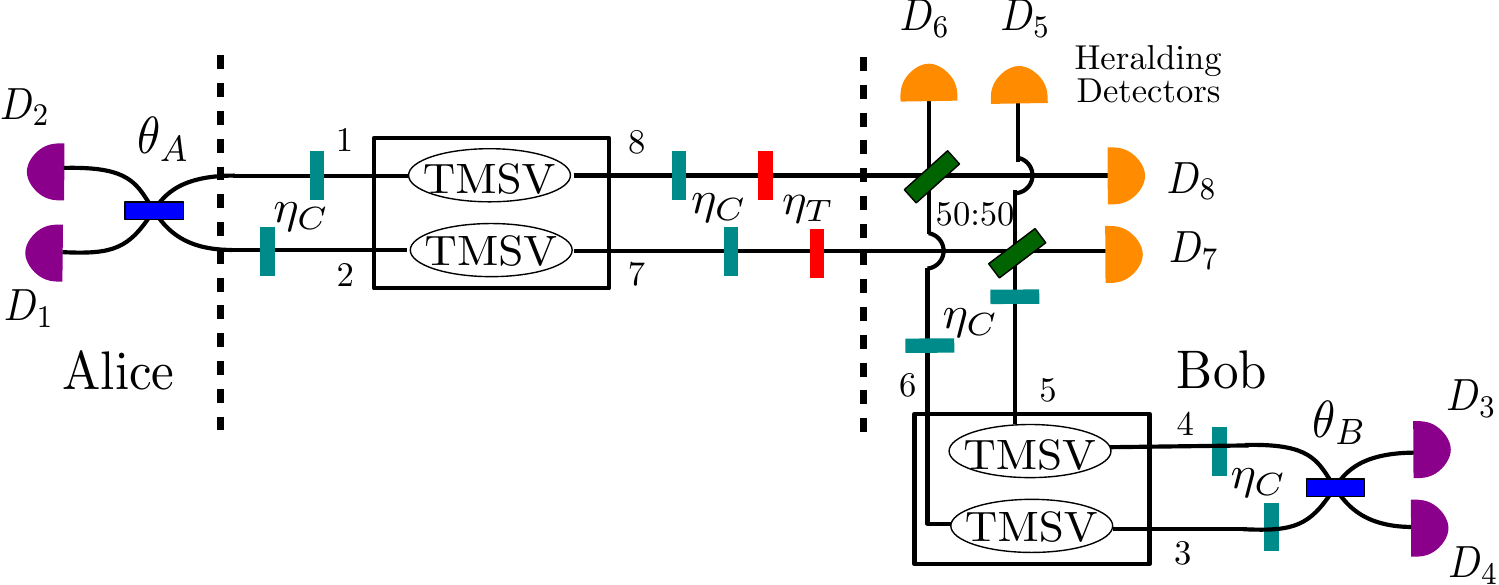}\caption{\label{fig:amp full pic} The ESR-assisted Bell testing setup for
DIQKD in greater detail. The Type-II SPDC crystal placed in a Sagnac
configuration as depicted in Fig. \ref{fig:source} is used to generate
a pair of two-mode squeezed vacuua (TMSV) over polarization modes.
The polarizer angle settings at Alice and Bob are denotes as $\theta_{A}$
and $\theta_{B}$, respectively. The detectors are assumed to be imperfect,
on-off photodetectors, with non-unit efficiencies and non-zero dark-count
probabilities.}
\end{figure*}

\subsection{Multimode spectral structure of the SPDC outputs}

When the parametric down conversion source is pumped by a pulsed laser,
the quantum state emitted has a spectral structure, and can be described
by \cite{GW97},
\begin{equation}
\exp\left[\xi\iint d\omega_{s}d\omega_{i}f(\omega_{s},\omega_{i})\hat{a}_{s}^{\dagger}(\omega_{s})\hat{a}_{i}^{\dagger}(\omega_{i})-h.c.\right]|0\rangle,\label{eq:emitted_state}
\end{equation}
where $\xi=r\exp\left(i\theta\right)$ is the squeezing parameter
(related to the pump power), $\hat{a}_{s}^{\dagger}(\omega_{s})$
and $\hat{a}_{i}^{\dagger}(\omega_{i})$ are creation operators for
the signal and idler modes with frequencies $\omega_{s}$ and $\omega_{i}$,
respectively. Let us call the above state $\left|\Psi\right\rangle $.
$f(\omega_{s},\omega_{i})$ is the joint spectral amplitude, which
is a product of the pump distribution and the phase-matching function
of the nonlinear crystal \cite{GW97,LWE00,MLSWUSW08}. We can assume
$\xi$ to be real and positive without losing generality.

This joint spectral amplitude can be decomposed using Schmidt decomposition
as \cite{LWE00,MLSWUSW08}
\begin{equation}
f(\omega_{s},\omega_{i})=\sum_{l}\sqrt{\lambda_{l}}g_{l}(\omega_{s})h_{l}(\omega_{i}),
\end{equation}
where $\lambda_{l}$, $g_{l}(\omega_{s})$, $h_{l}(\omega_{i})$ are
solutions of the eigenvalue equations:

\begin{eqnarray}
\intop K_{1}(\omega,\omega')g_{l}(\omega') & = & \lambda_{l}g_{l}(\omega'),\label{eq:eigenvalue_equation1}\\
\int K_{2}(\omega,\omega')h_{l}(\omega') & = & \lambda_{l}h_{l}(\omega'),
\end{eqnarray}
and
\begin{eqnarray}
K_{1}(\omega,\omega') & \equiv & \int d\omega_{2}f(\omega,\omega_{2})f^{*}(\omega',\omega_{2}),\label{eq:K}\\
K_{2}(\omega,\omega') & \equiv & \int d\omega_{1}f(\omega_{1},\omega)f^{*}(\omega_{1},\omega').
\end{eqnarray}
Then the state in (\ref{eq:emitted_state}) can be represented as
\begin{eqnarray}
|\Psi\rangle & = & \exp\left[r\sum_{l}\sqrt{\lambda_{l}}\hat{b}_{l}^{\dagger}\hat{c}_{l}^{\dagger}-h.c.\right]|0\rangle\label{eq:exp1}\\
 & = & \prod_{l}\exp\left[r\sqrt{\lambda_{l}}\hat{b}_{l}^{\dagger}\hat{c}_{l}^{\dagger}-h.c.\right]|0\rangle\label{eq:exp2}\\
 & = & |\Psi(r\sqrt{\lambda_{1}})\rangle|\Psi(r\sqrt{\lambda_{2}})\rangle\cdots,\label{eq: schmidt modes}
\end{eqnarray}
where
\begin{eqnarray}
\hat{b}_{l}^{\dagger} & =\int & d\omega_{s}g_{l}(\omega_{s})\hat{a}_{s}^{\dagger}(\omega_{s}),\label{eq:b,c}\\
\hat{c}_{l}^{\dagger} & =\int & d\omega_{i}h_{l}(\omega_{i})\hat{a}_{i}^{\dagger}(\omega_{i}),
\end{eqnarray}
and $\lambda_{l}$ is the Schmidt eigenvalue. Note that $\hat{b}_{l}$
and $\hat{c}_{l}$ satisfy a standard bosonic commutation relation
$[\hat{b}_{l},\hat{b}_{m}^{\dagger}]=[\hat{c}_{l},\hat{c}_{m}^{\dagger}]=\delta_{lm}$.
The decomposition of the exponential term as given in \eqref{eq:exp2}
is possible since the Schmidt modes are orthonormal. Finally, (\ref{eq: schmidt modes})
represents that in the Schmidt mode basis, the state is described
by tensor products of two-mode squeezed vacuums $|\Psi(r\sqrt{\lambda_{l}})\rangle$,
where 
\begin{equation}
|\Psi(r\sqrt{\lambda_{l}})\rangle=\frac{1}{\cosh r\sqrt{\lambda_{l}}}\sum_{n}\left(\tanh r\sqrt{\lambda_{l}}\right)^{n}|n\rangle_{B_{l}}|n\rangle_{C_{l}},\label{eq:Scmidt_TMSV}
\end{equation}
with the effective squeezing parameter $r\sqrt{\lambda_{l}}$. As
a consequence, we conclude that the quantum state emitted from the
SPDC source is simply given by a tensor product of two-mode squeezed
vacuums. Thus, in the Sagnac loop source-based described previously,
when the pump is a pulsed laser, the state in (\ref{eq:source-tmsv})
is further a tensor product of TMSVs over appropriate Schmidt modes.

Connection between $r$ and the experimentally observable parameter
is the following. Note that the theoretical modeling of $f(\omega_{s},\omega_{i})$
is well established and thus one can derive the Schmidt eigenvalues
for a given setup of the SPDC source. In the experiment, one can also
estimate the photon-pair generation rate of the SPDC source, i.e.
the probability that the source emits non-zero photons:
\begin{equation}
p=1-\prod_{l}\left|\langle00|\Psi(r\sqrt{\lambda_{l}})\rangle\right|^{2}.\label{eq:pair_gen_rate}
\end{equation}
Plugging (\ref{eq:Scmidt_TMSV}) into, (\ref{eq:pair_gen_rate}),
we obtain the relation
\begin{equation}
p=1-\prod_{l}\left(\cosh r\sqrt{\lambda_{l}}\right)^{-2},\label{eq:pair_gen_rate_vs_r}
\end{equation}
which allows us to derive $r$ numerically from experimentally estimated
$p$.

With the above observation and the recent theoretical method developed
in \cite{TJS15}, one can, e.g., simulate the four-photon HOM experiment
including experimental imperfections, infinitely higher order multi-photon
components, and joint spectral property of the SPDC source.

\section{Results \label{sec:Nonlocality-ESR}}

Having described our realistic models for the source and the detectors,
we now analyze the performance of the ESR-assisted DIQKD scheme with
realistic elements. We do so using the characteristic function-based
approach from Gaussian quantum information (see the appendix and \cite{TJS15}
for more details on the tools we use to perform our calculations).
The approach is quite effective to describe and analyze the system
consisting of Gaussian elements and on-off photodetectors, taking
into account multi-mode structure in the sources, and losses and dark
counts in the detectors.

Consider the full, linear optics-based depiction of the ESR-assisted
scheme for DIQKD shown in Fig. \ref{fig:amp full pic}. In this Figure,
for simplicity the modes are renumbered 1 through 8, with the odd-numbered
modes denoting horizontally polarized modes and the even-numbered
modes being vertically polarized. They are generated by SPDC as described
in Section \ref{sub:Polarization-entanglement} with the pairs 1 and
8, 2 and 7, etc., being in the two-mode squeezed vacuum state. The
polarizers are replaced by beamsplitters between the horizontal/vertical
mode pairs, with the tunable transmittivities $\cos^{2}\theta_{A}$
and $\cos^{2}\theta_{B}$ denoting the polarizer settings. The detectors
are assumed to be imperfect, on-off photodetectors, with non-unit
efficiencies and non-zero dark-count probabilities.

Firstly, we recall the results presented in \cite{CM11}, where Bob
employs the ESR node, but with the detectors modeled as PNRDs. A conclusive
detection event in this case refers to the presence of exactly a single
photon in the mode. Hence a conclusive-conclusive event at Alice and
Bob corresponds to the presence of a maximally entangled photon pair
with an intrinsic $S$ value of $2\sqrt{{2}}$. Any other combination
of events at Alice and Bob corresponds to the classical value of $S=2$.
So, the maximal possible violation could be written as the linear
combination $S=\mu_{cc}2\sqrt{{2}}+\left(1-\mu_{cc}\right)2$, where
$\mu_{cc}$ is the probability of obtaining a conclusive-conclusive
event at Alice and Bob. This, when evaluated in the limit of small
average photon numbers in the source, resulted in $S\approx1+\sqrt{{2}}$
, a constant independent of the distance of transmission. 

On the contrary, in our case where we consider on-off photodetectors,
a conclusive-conclusive event does not necessarily imply the presence
of a maximally entangled photon pair. Thus, we cannot adopt the analysis
of \cite{CM11} and are forced to resort to numerical optimization
to determine $S$ values for the state under different conditions
of the sources and the detectors. We globally optimize $S$ over the
measurement settings at Alice and Bob and the mean photon numbers
of the SPDC outputs at the two sources (one being the primary source
and the other at the relay node). We use a simulated annealing-based
numerical optimization algorithm. We find that the optimal measurement
angles at Alice and Bob for the considered loophole-free Bell test
are given by $\{0,\pi/6\}$ and $\left\{ \pi/2,2\pi/3\right\} $ for
the two parties, respectively, and are independent of all other conditions.
Assuming symmetric losses in orthogonal polarization modes, we further
optimize over an absolute mean photon number, the ratio between the
mean photon numbers of the primary source and the source in the relay
node, and the ratio between the mean photon numbers of the two SPDC
within a source. 

We present our results in two parts. Firstly, we focus on the potential
$S$ value of the state heralded upon successful entanglement swapping
as a function of the communication distance assuming a telecommunication
fiber of attenuation $\alpha=0.2$ dB/km. Here, we assume ideal coupling
and detectors at the end users Alice and Bob, but real, imperfect
ones at the relay node. Secondly, we consider real, imperfect coupling
and detectors all over including at Alice and Bob and analyze the
$S$ value as a function of distance. Here, we also separately optimize
$K$ to evaluate the performance of the scheme for key distribution.
We assume that the sources have a single pure Schmidt mode by default
unless mentioned otherwise. In the latter cases, we assume that there
are predominantly two Schmidt modes and we denote the leading Schmidt
eigenvalue as $\lambda$, with the other being $1-\lambda$.

\begin{figure}
\includegraphics[scale=0.4]{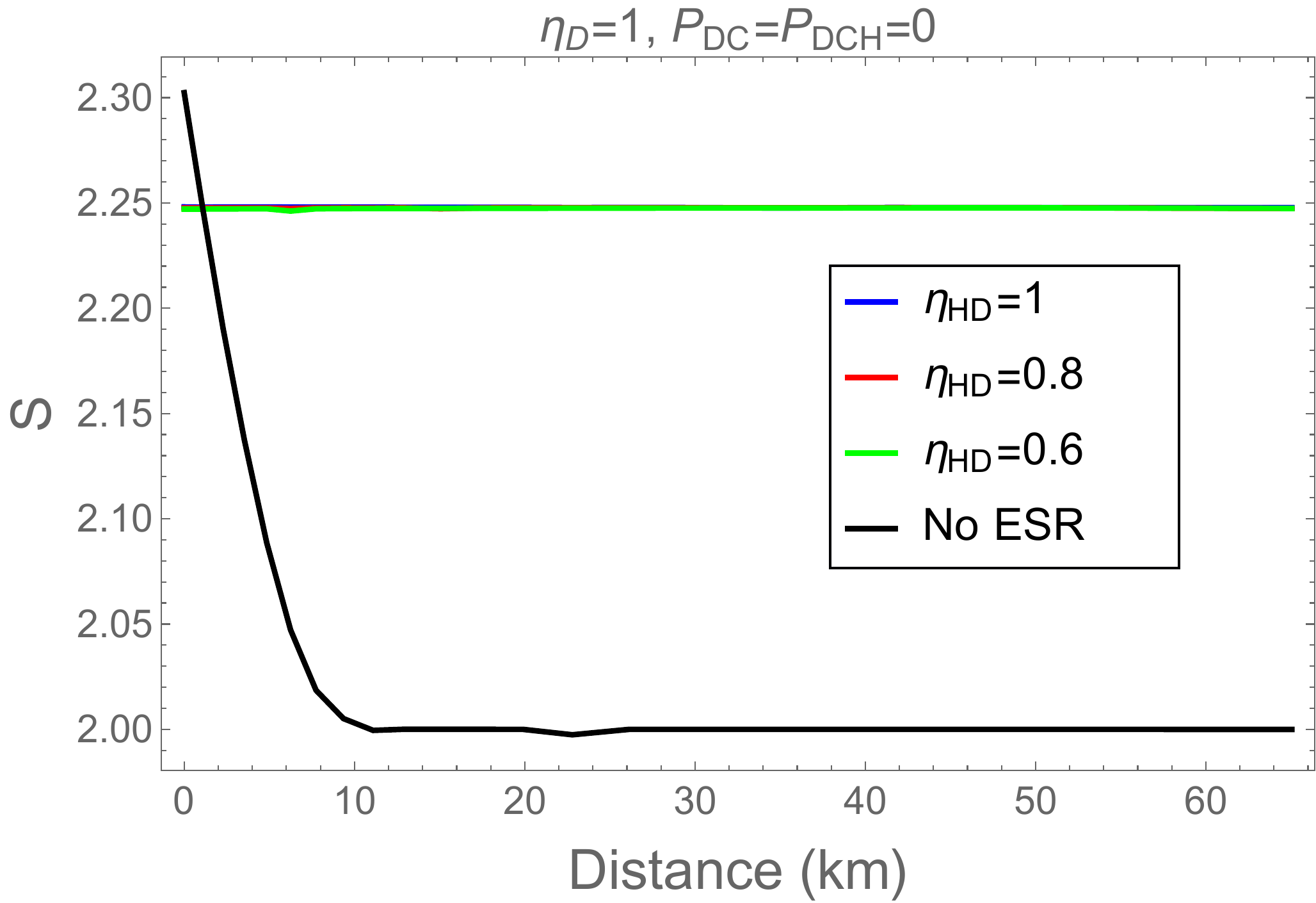}

(a)

\bigskip{}

\includegraphics[scale=0.4]{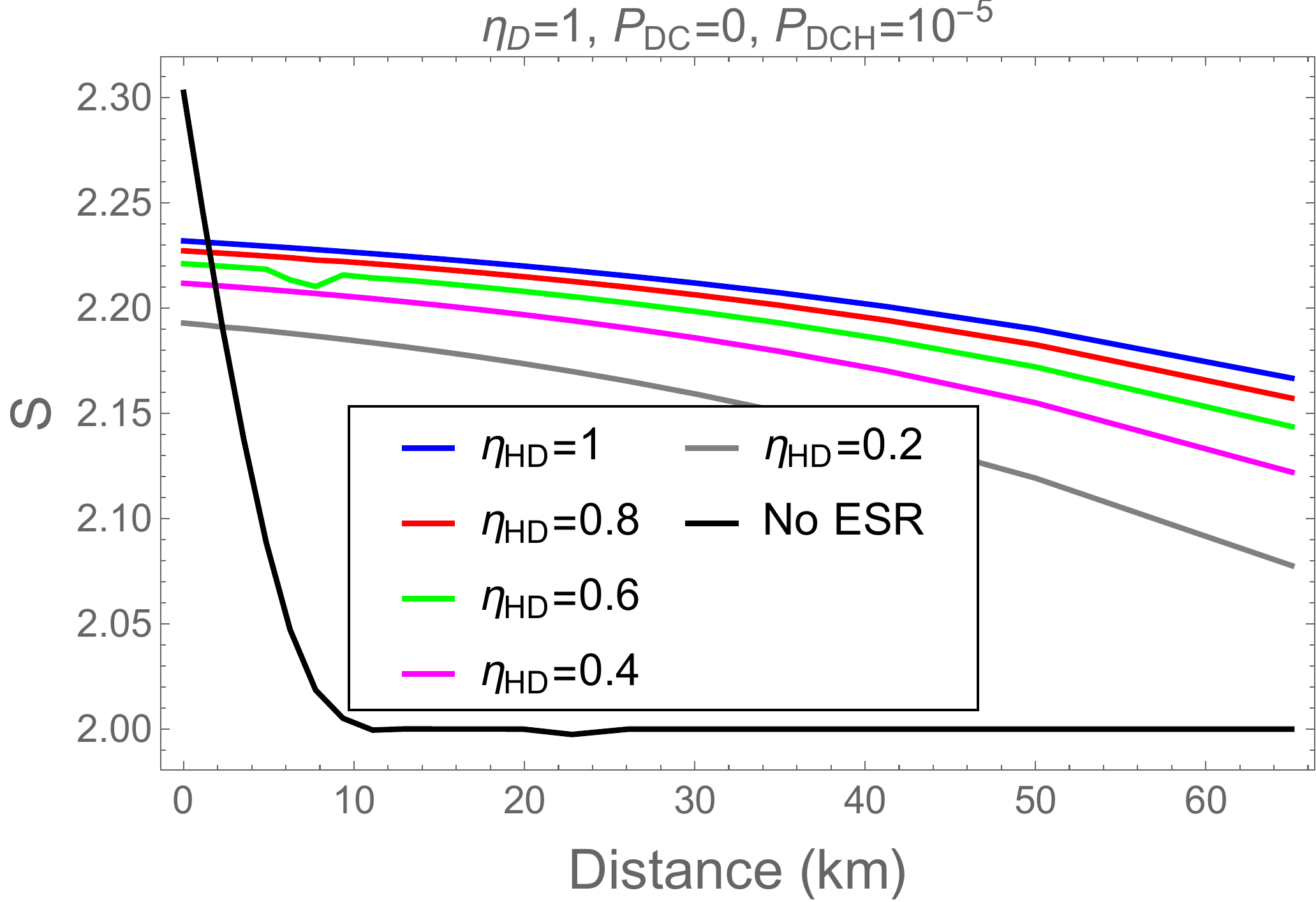}

(b)

\caption{\label{fig:S_idet}Maximal loophole-free violation of the CHSH inequality
$S$ as a function of distance in the ESR-assisted DIQKD of Fig. \ref{fig:amp full pic}.
The sources are assumed to be monochromatic. The detection efficiencies
at Alice and Bob are assumed to be ideal (i.e., the product of coupling
and detector efficiencies $\eta_{D}=\eta_{C}\eta_{\operatorname{det}}=1$
and dark count probability in the detectors $P_{DC}=0$). The quantities
$\eta_{HD}=\eta_{C}\eta_{\operatorname{hdet}}$ and $P_{DCH}$ denote
the detection efficiency and dark count probability, respectively,
for the heralding modes / detectors. Curves corresponding to various
values of $\eta_{DCH}$ are plotted for the cases (a) without ($P_{DCH}=0$)
and (b) with dark counts ($P_{DCH}\protect\neq0$) in the heralding
detectors. The black (reference) curve in both (a) and (b) corresponds
to the case where the ESR node is absent.}
\end{figure}

\begin{figure}
\includegraphics[scale=0.4]{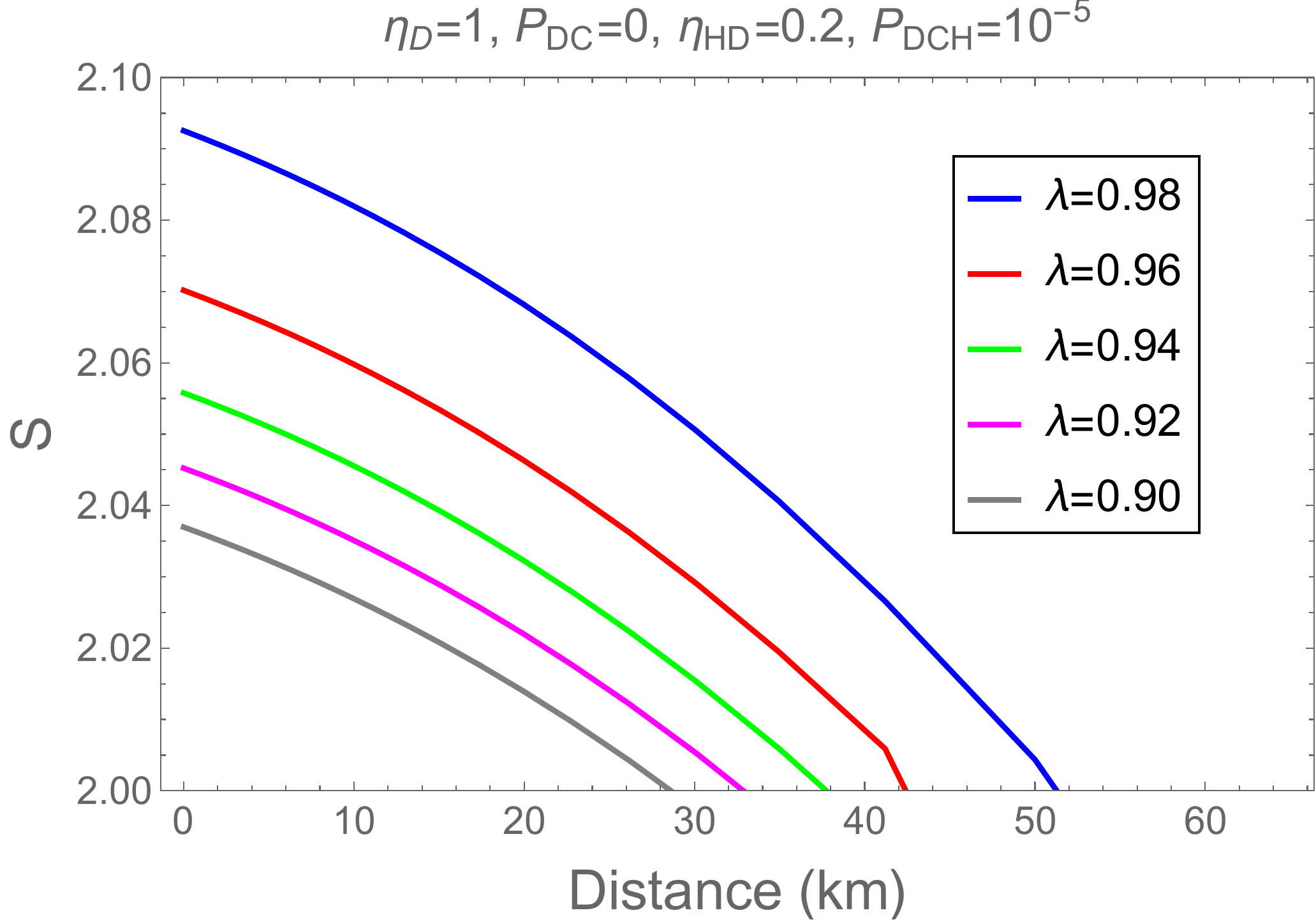}

\caption{\label{fig:S_idet_lambda_dc} Maximal loophole-free violation $S$
as a function of distance for different spectral spreads in the sources
for the ESR-assisted DIQKD scheme. $\lambda$ here corresponds to
the largest Schmidt eigenvalue in the Schmidt decomposition of the
joint spectral density. The detection efficiencies at Alice and Bob
are assumed to be ideal (i.e., the product of coupling and detector
efficiencies $\eta_{D}=\eta_{C}\eta_{\operatorname{det}}=1$ and dark
count probability in the detectors $P_{DC}=0$), while the detection
efficiencies in the heralding modes are taken as $\eta_{HD}=0.2$
and the dark count probability in the heralding detectors as $P_{DCH}=10^{-5}$.}
 
\end{figure}

\subsection{With ideal coupling and detectors at Alice and Bob, but real ones
at the relay node}

\begin{comment}
An analysis in \cite{VSBLC15} showed that a loophole-free CHSH test
based on the deterministic strategy described above in Section \ref{sec:ESR}
requires a minimum detection efficiency of 2/3 in order to exhibit
a violation of the inequality. Assuming ideal fiber coupling and detectors,
this corresponds to a distance of $8.8$km for the optical fiber communication
channel (using $\alpha=0.2$dB/km as the fiber attenuation constant).
We plot this in Figs. \eqref{fig:S_idet} (a) and (b) for comparison,
to show the merit of using the ESR node.
\end{comment}
Assuming ideal detectors at Alice and Bob, we find that the $S$ value
for the heralded state is constant over distance and is independent
of the efficiency of the heralding detectors (Fig. \ref{fig:S_idet}
(a)). Although the value of $S$ is less than the maximum violation
at zero distance obtainable in the absence of the ESR (see black curve
in Fig. \ref{fig:S_idet}), the fact that it is independent of distance
in principle is interesting. As explained previously, this feature
was also observed in \cite{CM11}, where PNRDs were employed. On the
other hand, when heralding detectors with a dark count probability
$P_{DCH}=10^{-5}$ are used, the constancy of $S$ no longer holds.
Nevertheless, its value is still significantly higher than the case
without the relay node for considerably larger range of distances
(Fig. \ref{fig:S_idet} (b)). 

Next, we include spectral spread in the sources, namely the multimode
nature $\lambda<1$. The detector efficiencies are assumed to be flat
over all the spectral modes. In Fig. \ref{fig:S_idet_lambda_dc},
we plot $S$ vs distance for various values of $\lambda$ for the
case of a realistic heralding detection efficiency of $\eta_{HD}=0.2$
and dark count probability $P_{DC}=10^{-5}$. We find that $S$ values
drops to the classical value of 2 faster with increasing spectral
impurity. The real cause for the faster degradation of $S$ vs distance
with increasing spectral impurity in the sources is that the on-off
detectors cannot discriminate clicks from different spectral modes,
which destroy the entanglement correlation in each mode. Thus, if
a Schmidt mode separator were possible to implement in front of the
detector set, there might be no degradation, but mode multiplexed
performance instead. Unfortunately such a separator is not easy to
realize. Nevertheless, the largest distance at which $S>2$ even for
$\lambda=0.95$ is about 30kms, which is still larger compared to
the 10km limit at which $S$ drops to 2 in the absence of the ESR
node.

\begin{figure}
\includegraphics[scale=0.4]{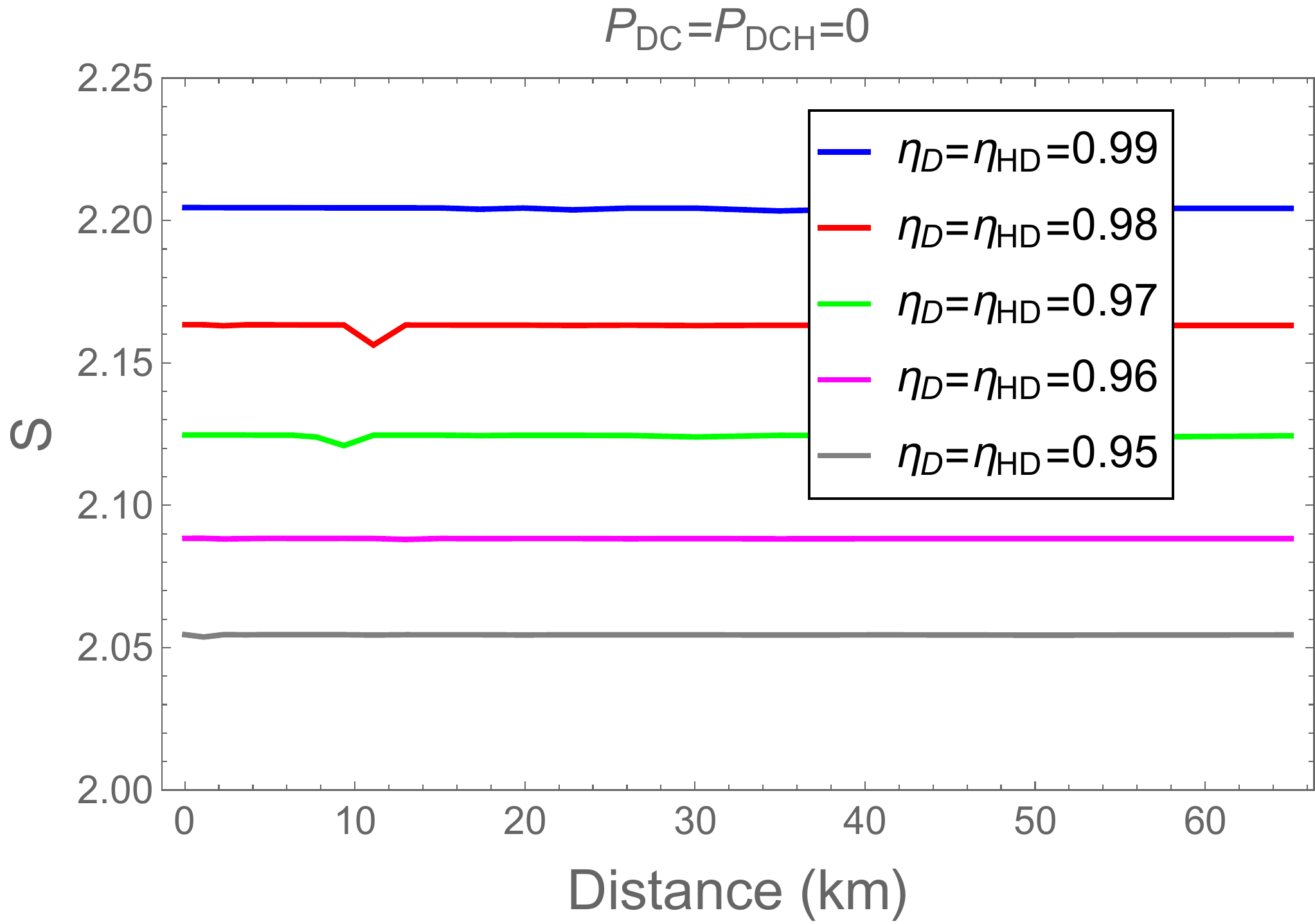}

(a)

\bigskip{}

\includegraphics[scale=0.4]{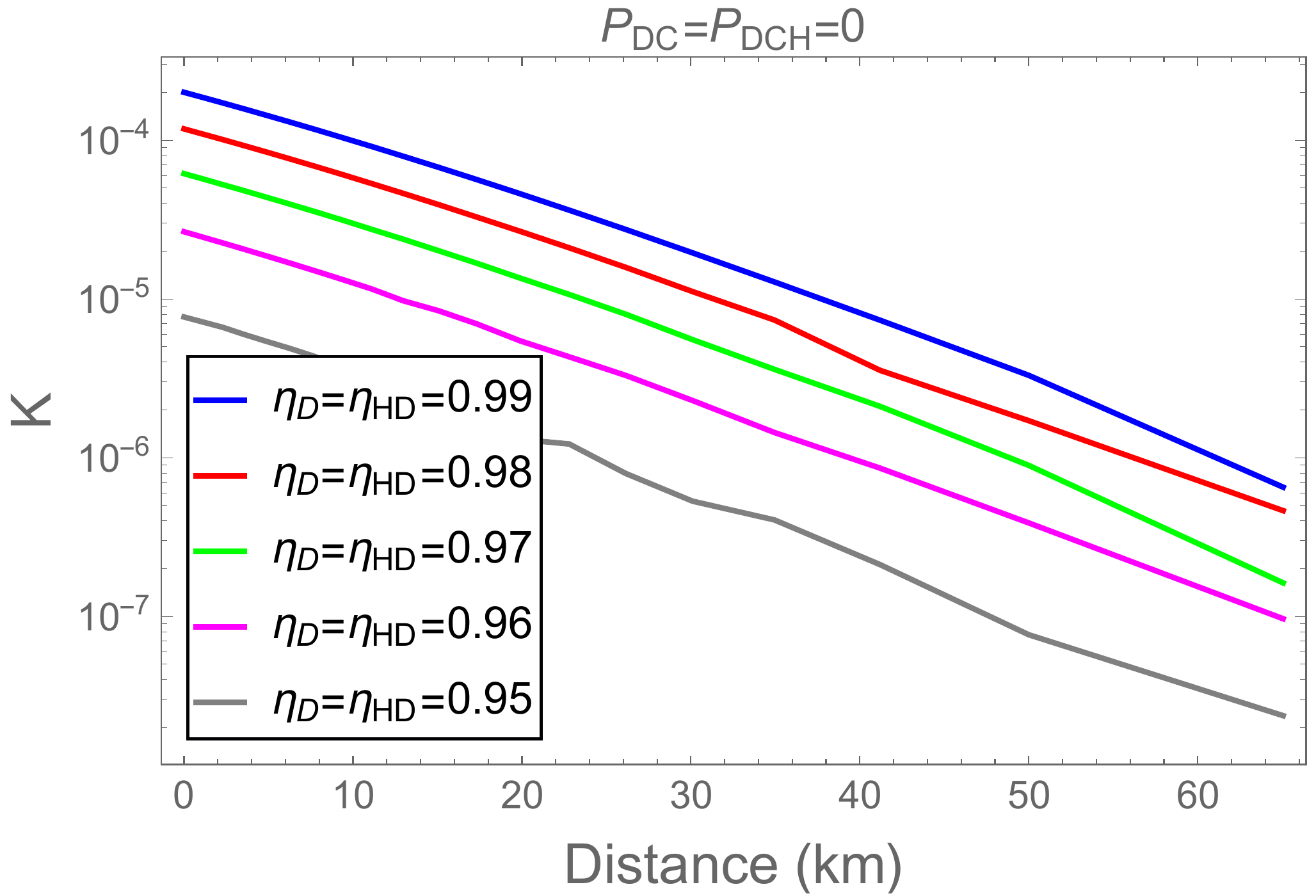}

(b)

\caption{\label{fig:rdet_nodc}(a) Maximal loophole-free violation of the CHSH
inequality $S$ and (b) a lower bound on the key rate $K$ (bits per
channel use), as a function of distance. The sources are assumed to
be monochromatic. All detection efficiencies are assumed to be of
non-unity ($\eta_{D}$ and $\eta_{HD}$ denoting the detection efficiencies
at the end users, and the heralding detectors, respectively), but
free from dark count ($P_{DC}=P_{DCH}=0$). Curves corresponding to
various values of detection efficiencies $\eta_{D}=\eta_{HD}=\eta$
are plotted.}
 
\end{figure}

\begin{figure}
\includegraphics[scale=0.4]{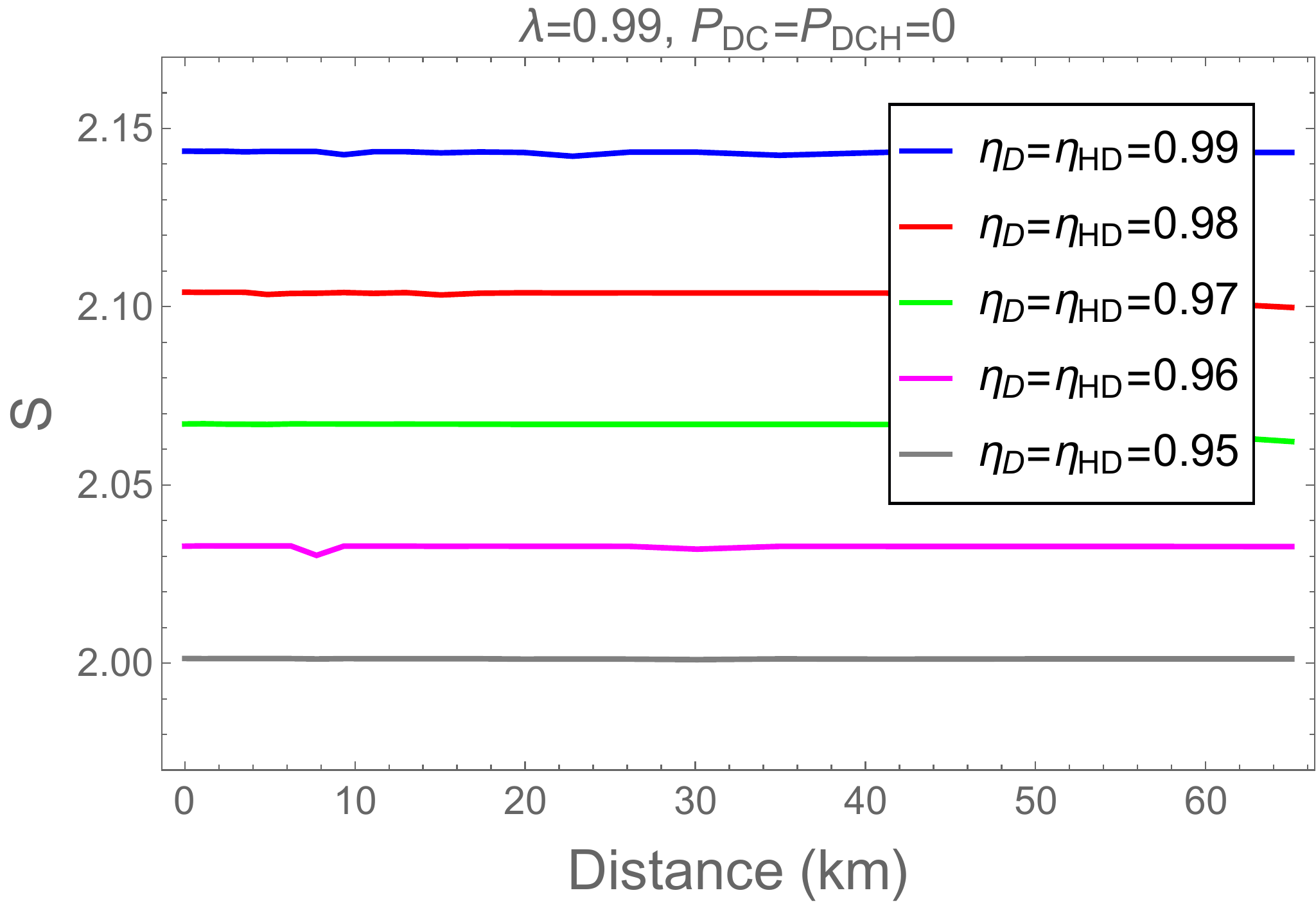}

(a)

\bigskip{}

\includegraphics[scale=0.4]{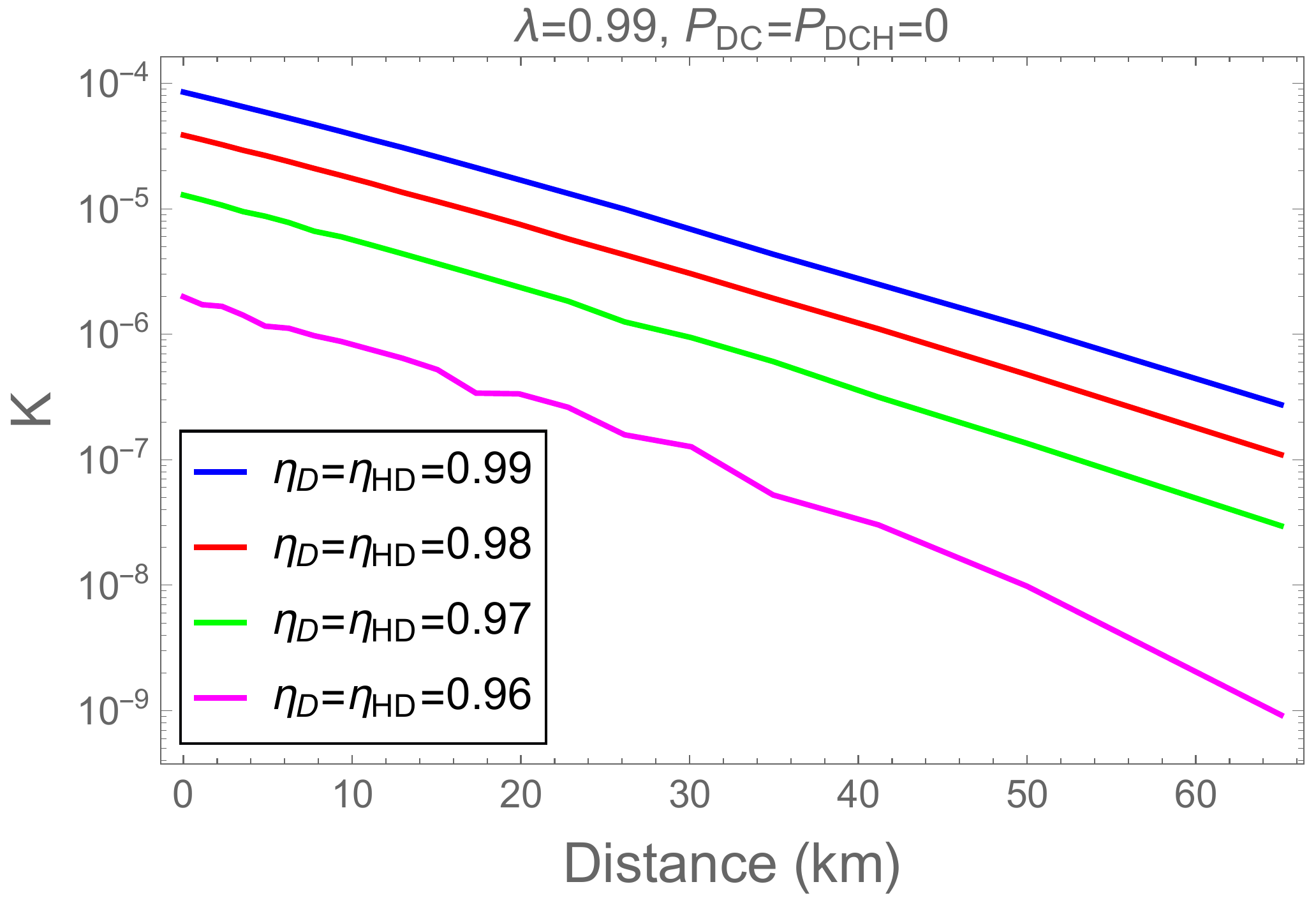}

(b)

\caption{\label{fig:rdet_lambda_nodc} Effect of spectral impurity (where the
leading Schmidt eigenvalue $\lambda$ is $0.99$) in the sources.
(a) The maximal loophole-free violation of the CHSH inequality $S$
and (b) a lower bound on the key rate, $K$ (bits per channel use),
are plotted as a function of distance. All detection efficiencies
are assumed to be of non-unity ($\eta_{D}$ and $\eta_{HD}$ denoting
the detection efficiencies at the end users, and the heralding detectors,
respectively), but free from dark count ($P_{DC}=P_{DCH}=0$). Curves
corresponding to various values of detection efficiencies $\eta_{D}=\eta_{HD}=\eta$
are plotted.}
 
\end{figure}

\begin{figure}
\includegraphics[scale=0.4]{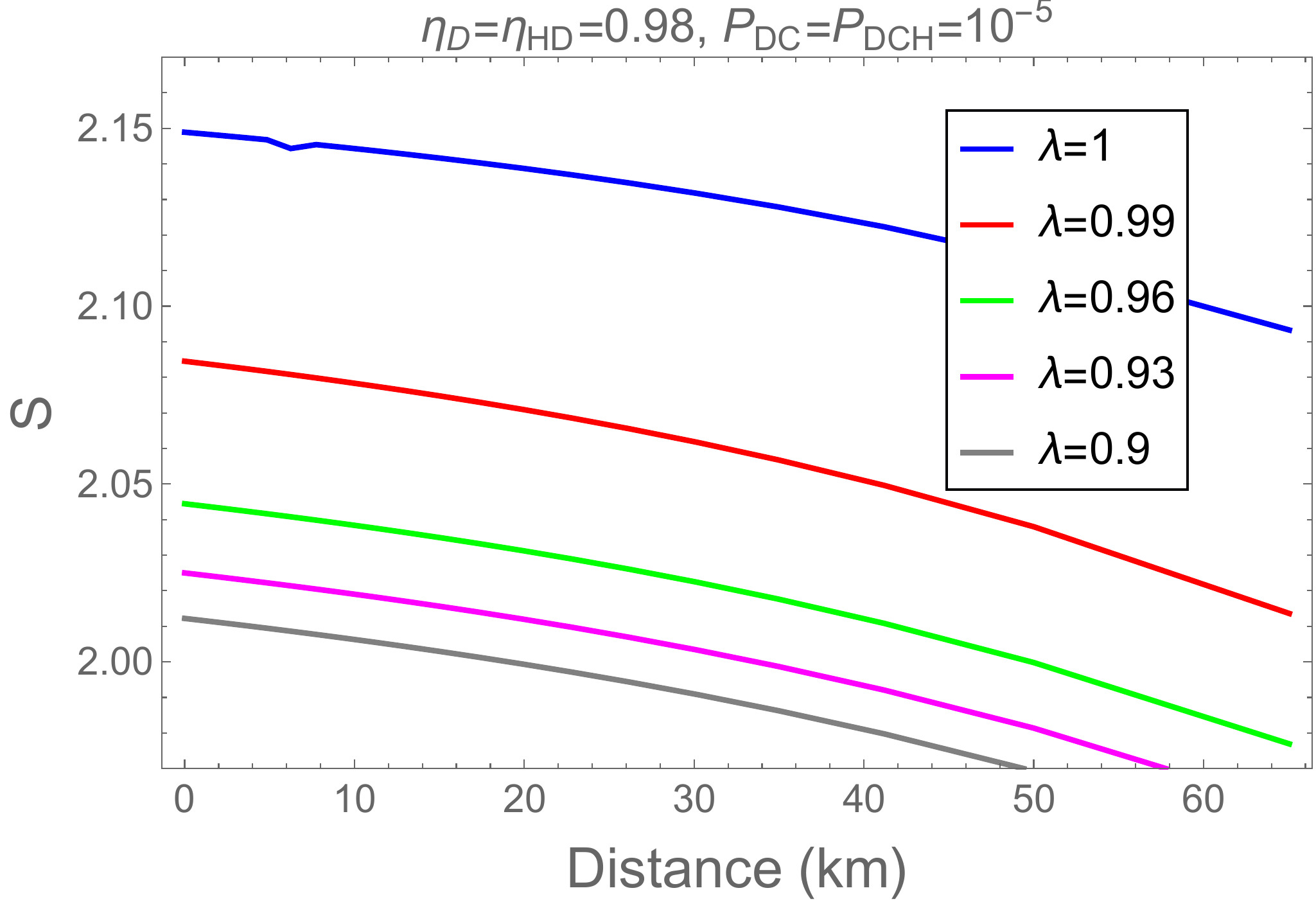}

(a)

\bigskip{}

\includegraphics[scale=0.38]{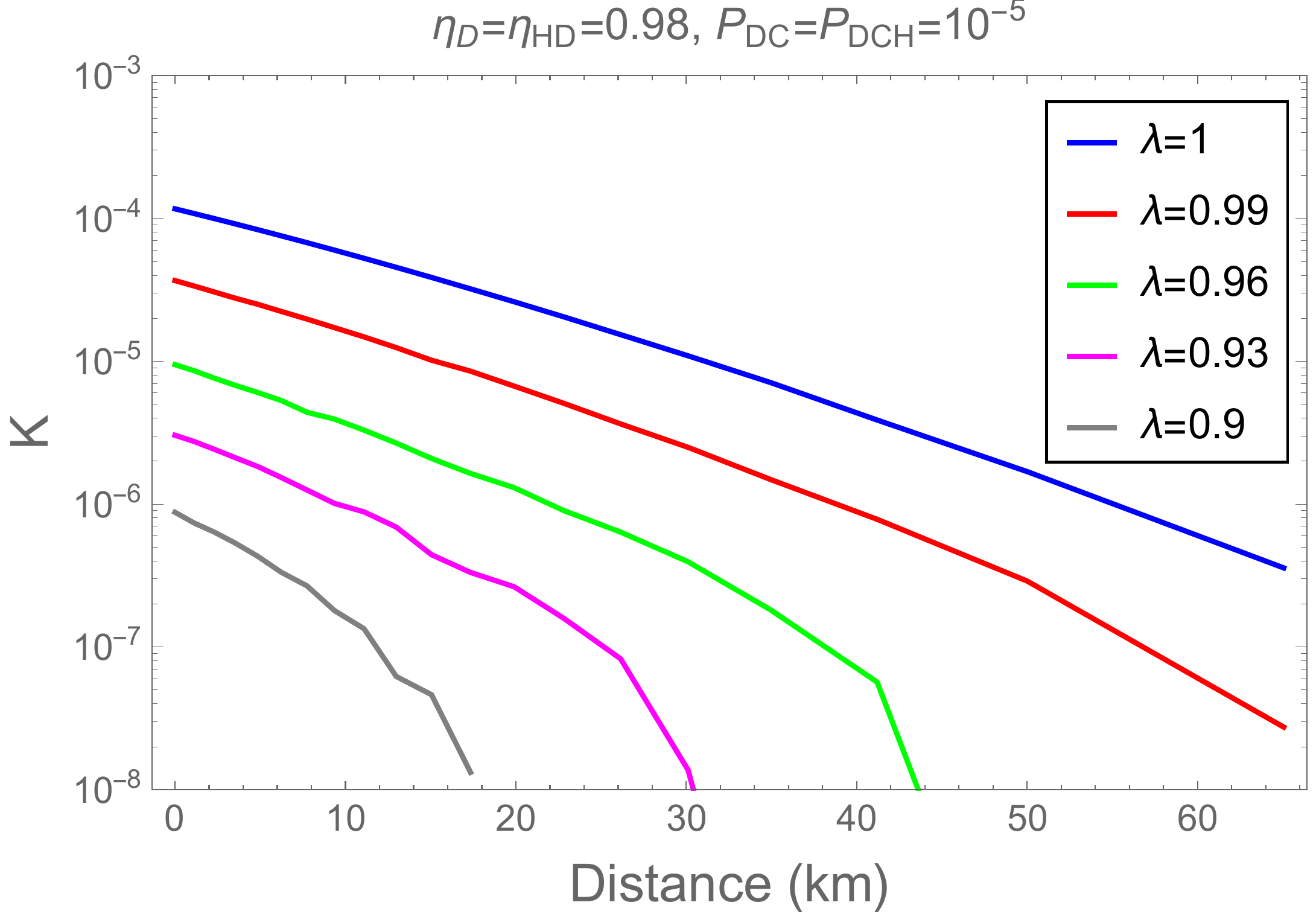}

(b)

\caption{\label{fig:rdet_lambda_dc}Variation due to the multimode spectral
structure in the sources. (a) The maximal loophole-free violation
of the CHSH inequality $S$ and (b) a lower bound on the key rate,
$K$ (bits per channel use), are plotted as a function of distance.
All detection efficiencies are assumed to be non-unity ($\eta_{D}=\eta_{HD}=0.98$,
where $\eta_{D}$ and $\eta_{HD}$ denote the detection efficiencies
at the end users, and the heralding detectors, respectively) and dark
count probability $P_{\operatorname{DC}}=10^{-5}$. The curves correspond
to different values of the largest Schmidt eigenvalue $\lambda$.}
 
\end{figure}

\subsection{With real, imperfect coupling and detectors all over}

Assuming identical, imperfect detectors at both Alice and Bob as well
as at the ESR, we now analyze the $S$ and $K$ values as functions
of distance. In the absence of dark counts in the detectors, we once
again find $S$ to be independent of the distance, but to decrease
towards the classical value of 2 with decreasing values of the detection
efficiencies $\eta_{D}=\eta_{HD}$ (Fig. \ref{fig:rdet_nodc} (a)).
The key rate $K$ similarly monotonically decreases with the detection
inefficiencies while keeping an exponential behavior in the drop with
respect to distance (Fig. \ref{fig:rdet_nodc} (b)). We now briefly
compare the $K$ values of Fig. \ref{fig:rdet_nodc} (b) with those
reported in \cite{CM11}. The calculations employed in the two are
identical, but the results are marginally different, because we use
on-off photodetector while \cite{CM11} used PNRDs. The PNRDs enable
higher $K$ values at short distances than the on-off detectors, but
we find that the performances of the two types of detectors even out
at larger distances. For, example, $K\approx10^{-6}$ bits per channel
use at a distance of about 60kms with both types of detectors. In
any case, the main point of emphasis here is that rates enabled by
the ESR are obviously much higher than what is possible without the
relay. 

Next, we include spectral imperfections in the sources. The same behavior
as above holds, but with diminished values of $S$ and $K$. Figs.
\ref{fig:rdet_lambda_nodc} (a) and (b) illustrate the point for sources
with a leading Schmidt eigenvalue of $\lambda=0.99$. Finally, when
dark counts are included ($P_{DC}=10^{-5}$), both $S$ and $K$ drop
with distance at faster rates corresponding to decreasing values of
$\lambda$ (Figs. \ref{fig:rdet_lambda_dc} (a) and (b)). The $K$
curves in this case exhibit the familiar cliff-type drop to zero when
the dark count rate becomes comparable to the signal rate, which decreases
with distance.

\section{Conclusions\label{sec:Discussion-and-Conclusion}}

We investigated a scheme for DIQKD that is based on the use of a simple,
conventional ESR node to mitigate the effect of transmission losses.
Going beyond earlier work of Curty and Moroder \cite{CM11}, we considered
a more realistic model for the entangled source and the detectors.
Our sources of polarization-entanglement were taken to be based on
a pair of pulsed SPDCs, having infinitely higher order multi-photon
components and multimode spectral structure. Our detectors were taken
to be spectrally-flat on-off photodetectors, which simply distinguish
the event of presence of photons from absence. The detectors included
losses (contributions from detector inefficiency and free-space to
fiber coupling inefficiency) and dark counts. We presented an exact
key rate analysis for the scheme based on the use of tools from Gaussian
quantum information. Our results showed that the relay node enables
positive key rates over larger distances than what is possible without
the relay node for sufficiently large detection efficiencies (which
includes detector and free-space to fiber coupling efficiencies),
small dark count probabilities in the detectors and small spectral
spread in the sources. Thus, our results established the robustness
of the ESR-based scheme for DIQKD against imperfections in the sources
and detectors.

While our analyses captured the effects of imperfections in the SPDC
sources and in the detection process to a large extent, there is room
for further refinement. For example, in the multimode spectral modeling
of the sources, more terms in the Schmidt decomposition could be included.
In the model for the on-off photodetectors, spectrally-dependent efficiencies
could be considered. Since the source is based on a pulsed laser,
temporal mode mismatch could be included in the overall model.

To conclude, our results ascertain that it is possible to mitigate
transmission losses using the ESR node with more realistic models
for the sources and detectors than what was considered in \cite{CM11}.
However, the ultimate practical realizability of DIQKD still hinges
on improvements the detector technologies. As noted in \cite{CM11}
and concurred by our analyses in this paper, detection efficiencies
upwards of 95\% are required to realize DIQKD even in the case that
the source spectral purity is one and the detectors are dark count-free.
Recent progress in coupling and detector technologies shows promise
that such high detection efficiencies might be achievable in the not-so-distant
future.
\begin{acknowledgments}
KPS thanks the National Institute of Information and Communications
Technologies, Tokyo, for their hospitality during the summer
of 2015, when a majority of this work was carried out. KPS acknowledges
funding from the National Science Foundation (NSF) under Award No. CCF-1350397 and the Max Planck Society. The authors thank Marcos Curty, Jonathan Dowling, Ruibo-Jin,
George Knee, Bill Munro, Kae Nemoto and Mark M. Wilde for valuable
discussions. This work was supported from Open Partnership Bilateral
Joint Research Projects (JSPS) and ImPACT Program of Council for Science,
Technology and Innovation (Cabinet Office, Government of Japan). 
\bibliographystyle{unsrt}
\bibliography{Ref}

\begin{thebibliography}{37}
\expandafter\ifx\csname natexlab\endcsname\relax\def\natexlab#1{#1}\fi
\expandafter\ifx\csname bibnamefont\endcsname\relax
  \def\bibnamefont#1{#1}\fi
\expandafter\ifx\csname bibfnamefont\endcsname\relax
  \def\bibfnamefont#1{#1}\fi
\expandafter\ifx\csname citenamefont\endcsname\relax
  \def\citenamefont#1{#1}\fi
\expandafter\ifx\csname url\endcsname\relax
  \def\url#1{\texttt{#1}}\fi
\expandafter\ifx\csname urlprefix\endcsname\relax\def\urlprefix{URL }\fi
\providecommand{\bibinfo}[2]{#2}
\providecommand{\eprint}[2][]{\url{#2}}

\bibitem[{\citenamefont{Scarani et~al.}(2009)\citenamefont{Scarani,
  Bechmann-Pasquinucci, Cerf, Du\ifmmode~\check{s}\else \v{s}\fi{}ek,
  L\"utkenhaus, and Peev}}]{SBCDLP09}
\bibinfo{author}{\bibfnamefont{V.}~\bibnamefont{Scarani}},
  \bibinfo{author}{\bibfnamefont{H.}~\bibnamefont{Bechmann-Pasquinucci}},
  \bibinfo{author}{\bibfnamefont{N.~J.} \bibnamefont{Cerf}},
  \bibinfo{author}{\bibfnamefont{M.}~\bibnamefont{Du\ifmmode~\check{s}\else
  \v{s}\fi{}ek}},
  \bibinfo{author}{\bibfnamefont{N.}~\bibnamefont{L\"utkenhaus}},
  \bibnamefont{and} \bibinfo{author}{\bibfnamefont{M.}~\bibnamefont{Peev}},
  \bibinfo{journal}{Reviews of Modern Physics} \textbf{\bibinfo{volume}{81}},
  \bibinfo{pages}{1301} (\bibinfo{year}{2009}),
  \bibinfo{note}{arXiv:0802.4155},
  \urlprefix\url{http://link.aps.org/doi/10.1103/RevModPhys.81.1301}.

\bibitem[{\citenamefont{Gisin et~al.}(2002)\citenamefont{Gisin, Ribordy,
  Tittel, and Zbinden}}]{GRTZ02}
\bibinfo{author}{\bibfnamefont{N.}~\bibnamefont{Gisin}},
  \bibinfo{author}{\bibfnamefont{G.}~\bibnamefont{Ribordy}},
  \bibinfo{author}{\bibfnamefont{W.}~\bibnamefont{Tittel}}, \bibnamefont{and}
  \bibinfo{author}{\bibfnamefont{H.}~\bibnamefont{Zbinden}},
  \bibinfo{journal}{Rev. Mod. Phys.} \textbf{\bibinfo{volume}{74}},
  \bibinfo{pages}{145} (\bibinfo{year}{2002}),
  \urlprefix\url{http://link.aps.org/doi/10.1103/RevModPhys.74.145}.

\bibitem[{\citenamefont{Bennett and Brassard}(1984)}]{BB84}
\bibinfo{author}{\bibfnamefont{C.~H.} \bibnamefont{Bennett}} \bibnamefont{and}
  \bibinfo{author}{\bibfnamefont{G.}~\bibnamefont{Brassard}}, in
  \emph{\bibinfo{booktitle}{Proceedings of IEEE International Conference on
  Computers Systems and Signal Processing}} (\bibinfo{address}{Bangalore,
  India}, \bibinfo{year}{1984}), pp. \bibinfo{pages}{175--179}.

\bibitem[{\citenamefont{Ekert}(1991)}]{E91}
\bibinfo{author}{\bibfnamefont{A.~K.} \bibnamefont{Ekert}},
  \bibinfo{journal}{Phys. Rev. Lett.} \textbf{\bibinfo{volume}{67}},
  \bibinfo{pages}{661} (\bibinfo{year}{1991}),
  \urlprefix\url{http://link.aps.org/doi/10.1103/PhysRevLett.67.661}.

\bibitem[{\citenamefont{Bennett}(1992)}]{B92}
\bibinfo{author}{\bibfnamefont{C.~H.} \bibnamefont{Bennett}},
  \bibinfo{journal}{Phys. Rev. Lett.} \textbf{\bibinfo{volume}{68}},
  \bibinfo{pages}{3121} (\bibinfo{year}{1992}),
  \urlprefix\url{http://link.aps.org/doi/10.1103/PhysRevLett.68.3121}.

\bibitem[{\citenamefont{Grosshans and Grangier}(2002)}]{GG02}
\bibinfo{author}{\bibfnamefont{F.}~\bibnamefont{Grosshans}} \bibnamefont{and}
  \bibinfo{author}{\bibfnamefont{P.}~\bibnamefont{Grangier}},
  \bibinfo{journal}{Phys. Rev. Lett.} \textbf{\bibinfo{volume}{88}},
  \bibinfo{pages}{057902} (\bibinfo{year}{2002}),
  \urlprefix\url{http://link.aps.org/doi/10.1103/PhysRevLett.88.057902}.

\bibitem[{\citenamefont{Scarani et~al.}(2004)\citenamefont{Scarani, Ac\'{\i}n,
  Ribordy, and Gisin}}]{SARG04}
\bibinfo{author}{\bibfnamefont{V.}~\bibnamefont{Scarani}},
  \bibinfo{author}{\bibfnamefont{A.}~\bibnamefont{Ac\'{\i}n}},
  \bibinfo{author}{\bibfnamefont{G.}~\bibnamefont{Ribordy}}, \bibnamefont{and}
  \bibinfo{author}{\bibfnamefont{N.}~\bibnamefont{Gisin}},
  \bibinfo{journal}{Phys. Rev. Lett.} \textbf{\bibinfo{volume}{92}},
  \bibinfo{pages}{057901} (\bibinfo{year}{2004}),
  \urlprefix\url{http://link.aps.org/doi/10.1103/PhysRevLett.92.057901}.

\bibitem[{\citenamefont{Lo et~al.}(2005)\citenamefont{Lo, Ma, and
  Chen}}]{LMC05}
\bibinfo{author}{\bibfnamefont{H.-K.} \bibnamefont{Lo}},
  \bibinfo{author}{\bibfnamefont{X.}~\bibnamefont{Ma}}, \bibnamefont{and}
  \bibinfo{author}{\bibfnamefont{K.}~\bibnamefont{Chen}},
  \bibinfo{journal}{Phys. Rev. Lett.} \textbf{\bibinfo{volume}{94}},
  \bibinfo{pages}{230504} (\bibinfo{year}{2005}),
  \urlprefix\url{http://link.aps.org/doi/10.1103/PhysRevLett.94.230504}.

\bibitem[{\citenamefont{Mayers and Yao}(1998)}]{MY98}
\bibinfo{author}{\bibfnamefont{D.}~\bibnamefont{Mayers}} \bibnamefont{and}
  \bibinfo{author}{\bibfnamefont{A.}~\bibnamefont{Yao}}, in
  \emph{\bibinfo{booktitle}{Proceedings of the 39th Annual Symposium on
  Foundations of Computer Science}} (\bibinfo{publisher}{IEEE Computer
  Society}, \bibinfo{address}{Washington, DC, USA}, \bibinfo{year}{1998}), FOCS
  '98, pp. \bibinfo{pages}{503--}, ISBN \bibinfo{isbn}{0-8186-9172-7},
  \urlprefix\url{http://dl.acm.org/citation.cfm?id=795664.796390}.

\bibitem[{\citenamefont{Vazirani and Vidick}(2014)}]{VV14}
\bibinfo{author}{\bibfnamefont{U.}~\bibnamefont{Vazirani}} \bibnamefont{and}
  \bibinfo{author}{\bibfnamefont{T.}~\bibnamefont{Vidick}},
  \bibinfo{journal}{Phys. Rev. Lett.} \textbf{\bibinfo{volume}{113}},
  \bibinfo{pages}{140501} (\bibinfo{year}{2014}),
  \urlprefix\url{http://link.aps.org/doi/10.1103/PhysRevLett.113.140501}.

\bibitem[{\citenamefont{Pironio et~al.}(2009)\citenamefont{Pironio, Acin,
  Brunner, Gisin, Massar, and Scarani}}]{PABGMS09}
\bibinfo{author}{\bibfnamefont{S.}~\bibnamefont{Pironio}},
  \bibinfo{author}{\bibfnamefont{A.}~\bibnamefont{Acin}},
  \bibinfo{author}{\bibfnamefont{N.}~\bibnamefont{Brunner}},
  \bibinfo{author}{\bibfnamefont{N.}~\bibnamefont{Gisin}},
  \bibinfo{author}{\bibfnamefont{S.}~\bibnamefont{Massar}}, \bibnamefont{and}
  \bibinfo{author}{\bibfnamefont{V.}~\bibnamefont{Scarani}},
  \bibinfo{journal}{New Journal of Physics} \textbf{\bibinfo{volume}{11}},
  \bibinfo{pages}{045021} (\bibinfo{year}{2009}),
  \urlprefix\url{http://stacks.iop.org/1367-2630/11/i=4/a=045021}.

\bibitem[{\citenamefont{Acin et~al.}(2007)\citenamefont{Acin, Brunner, Gisin,
  Massar, Pironio, and Scarani}}]{ABGMPS07}
\bibinfo{author}{\bibfnamefont{A.}~\bibnamefont{Acin}},
  \bibinfo{author}{\bibfnamefont{N.}~\bibnamefont{Brunner}},
  \bibinfo{author}{\bibfnamefont{N.}~\bibnamefont{Gisin}},
  \bibinfo{author}{\bibfnamefont{S.}~\bibnamefont{Massar}},
  \bibinfo{author}{\bibfnamefont{S.}~\bibnamefont{Pironio}}, \bibnamefont{and}
  \bibinfo{author}{\bibfnamefont{V.}~\bibnamefont{Scarani}},
  \bibinfo{journal}{Phys. Rev. Lett.} \textbf{\bibinfo{volume}{98}},
  \bibinfo{pages}{230501} (\bibinfo{year}{2007}),
  \urlprefix\url{http://link.aps.org/doi/10.1103/PhysRevLett.98.230501}.

\bibitem[{\citenamefont{Brunner et~al.}(2014)\citenamefont{Brunner, Cavalcanti,
  Pironio, Scarani, and Wehner}}]{BCPSW14}
\bibinfo{author}{\bibfnamefont{N.}~\bibnamefont{Brunner}},
  \bibinfo{author}{\bibfnamefont{D.}~\bibnamefont{Cavalcanti}},
  \bibinfo{author}{\bibfnamefont{S.}~\bibnamefont{Pironio}},
  \bibinfo{author}{\bibfnamefont{V.}~\bibnamefont{Scarani}}, \bibnamefont{and}
  \bibinfo{author}{\bibfnamefont{S.}~\bibnamefont{Wehner}},
  \bibinfo{journal}{Rev. Mod. Phys.} \textbf{\bibinfo{volume}{86}},
  \bibinfo{pages}{419} (\bibinfo{year}{2014}),
  \urlprefix\url{http://link.aps.org/doi/10.1103/RevModPhys.86.419}.

\bibitem[{\citenamefont{Hensen et~al.}(2015)\citenamefont{Hensen, Bernien,
  Dreau, Reiserer, Kalb, Blok, Ruitenberg, Vermeulen, Schouten, Abellan
  et~al.}}]{Henson15}
\bibinfo{author}{\bibfnamefont{B.}~\bibnamefont{Hensen}},
  \bibinfo{author}{\bibfnamefont{H.}~\bibnamefont{Bernien}},
  \bibinfo{author}{\bibfnamefont{A.~E.} \bibnamefont{Dreau}},
  \bibinfo{author}{\bibfnamefont{A.}~\bibnamefont{Reiserer}},
  \bibinfo{author}{\bibfnamefont{N.}~\bibnamefont{Kalb}},
  \bibinfo{author}{\bibfnamefont{M.~S.} \bibnamefont{Blok}},
  \bibinfo{author}{\bibfnamefont{J.}~\bibnamefont{Ruitenberg}},
  \bibinfo{author}{\bibfnamefont{R.~F.~L.} \bibnamefont{Vermeulen}},
  \bibinfo{author}{\bibfnamefont{R.~N.} \bibnamefont{Schouten}},
  \bibinfo{author}{\bibfnamefont{C.}~\bibnamefont{Abellan}},
  \bibnamefont{et~al.}, \bibinfo{journal}{Nature}
  \textbf{\bibinfo{volume}{advance online publication}} (\bibinfo{year}{2015}),
  ISSN \bibinfo{issn}{1476-4687},
  \urlprefix\url{http://dx.doi.org/10.1038/nature15759 10.1038/nature15759
  http://www.nature.com/nature/journal/vaop/ncurrent/abs/nature15759.html\#supplementary-information}.

\bibitem[{\citenamefont{Giustina et~al.}(2015)\citenamefont{Giustina,
  Versteegh, Wengerowsky, Handsteiner, Hochrainer, Phelan, Steinlechner,
  Kofler, Larsson, Abellan et~al.}}]{GVW15}
\bibinfo{author}{\bibfnamefont{M.}~\bibnamefont{Giustina}},
  \bibinfo{author}{\bibfnamefont{M.~A.~M.} \bibnamefont{Versteegh}},
  \bibinfo{author}{\bibfnamefont{S.}~\bibnamefont{Wengerowsky}},
  \bibinfo{author}{\bibfnamefont{J.}~\bibnamefont{Handsteiner}},
  \bibinfo{author}{\bibfnamefont{A.}~\bibnamefont{Hochrainer}},
  \bibinfo{author}{\bibfnamefont{K.}~\bibnamefont{Phelan}},
  \bibinfo{author}{\bibfnamefont{F.}~\bibnamefont{Steinlechner}},
  \bibinfo{author}{\bibfnamefont{J.}~\bibnamefont{Kofler}},
  \bibinfo{author}{\bibfnamefont{J.-A.} \bibnamefont{Larsson}},
  \bibinfo{author}{\bibfnamefont{C.}~\bibnamefont{Abellan}},
  \bibnamefont{et~al.}, \emph{\bibinfo{title}{{A} significant-loophole-free
  test of {B}ell's theorem with entangled photons}} (\bibinfo{year}{2015}),
  \bibinfo{note}{arXiv:1511.03190v1}, \eprint{1511.03190}.

\bibitem[{\citenamefont{Shalm et~al.}(2015)\citenamefont{Shalm, Meyer-Scott,
  Christensen, Bierhorst, Wayne, Stevens, Gerrits, Glancy, Hamel, Allman
  et~al.}}]{SMC15}
\bibinfo{author}{\bibfnamefont{L.~K.} \bibnamefont{Shalm}},
  \bibinfo{author}{\bibfnamefont{E.}~\bibnamefont{Meyer-Scott}},
  \bibinfo{author}{\bibfnamefont{B.~G.} \bibnamefont{Christensen}},
  \bibinfo{author}{\bibfnamefont{P.}~\bibnamefont{Bierhorst}},
  \bibinfo{author}{\bibfnamefont{M.~A.} \bibnamefont{Wayne}},
  \bibinfo{author}{\bibfnamefont{M.~J.} \bibnamefont{Stevens}},
  \bibinfo{author}{\bibfnamefont{T.}~\bibnamefont{Gerrits}},
  \bibinfo{author}{\bibfnamefont{S.}~\bibnamefont{Glancy}},
  \bibinfo{author}{\bibfnamefont{D.~R.} \bibnamefont{Hamel}},
  \bibinfo{author}{\bibfnamefont{M.~S.} \bibnamefont{Allman}},
  \bibnamefont{et~al.}, \emph{\bibinfo{title}{{A} strong loophole-free test of
  local realism}} (\bibinfo{year}{2015}), \bibinfo{note}{arXiv:1511.03189v1},
  \eprint{1511.03189}.

\bibitem[{\citenamefont{Ralph and Lund}(2009)}]{RL09}
\bibinfo{author}{\bibfnamefont{T.~C.} \bibnamefont{Ralph}} \bibnamefont{and}
  \bibinfo{author}{\bibfnamefont{A.~P.} \bibnamefont{Lund}},
  \bibinfo{journal}{AIP Conference Proceedings}
  \textbf{\bibinfo{volume}{1110}}, \bibinfo{pages}{155} (\bibinfo{year}{2009}),
  \urlprefix\url{http://scitation.aip.org/content/aip/proceeding/aipcp/10.1063/1.3131295}.

\bibitem[{\citenamefont{Gisin et~al.}(2010)\citenamefont{Gisin, Pironio, and
  Sangouard}}]{GPS10}
\bibinfo{author}{\bibfnamefont{N.}~\bibnamefont{Gisin}},
  \bibinfo{author}{\bibfnamefont{S.}~\bibnamefont{Pironio}}, \bibnamefont{and}
  \bibinfo{author}{\bibfnamefont{N.}~\bibnamefont{Sangouard}},
  \bibinfo{journal}{Phys. Rev. Lett.} \textbf{\bibinfo{volume}{105}},
  \bibinfo{pages}{070501} (\bibinfo{year}{2010}),
  \urlprefix\url{http://link.aps.org/doi/10.1103/PhysRevLett.105.070501}.

\bibitem[{\citenamefont{Lvovsky}(2013)}]{Lvov13}
\bibinfo{author}{\bibfnamefont{A.~I.} \bibnamefont{Lvovsky}},
  \bibinfo{journal}{Nat Phys} \textbf{\bibinfo{volume}{9}}, \bibinfo{pages}{5}
  (\bibinfo{year}{2013}), ISSN \bibinfo{issn}{1745-2473},
  \urlprefix\url{http://dx.doi.org/10.1038/nphys2517}.

\bibitem[{\citenamefont{Kocsis et~al.}(2013)\citenamefont{Kocsis, Xiang, Ralph,
  and Pryde}}]{KXRP13}
\bibinfo{author}{\bibfnamefont{S.}~\bibnamefont{Kocsis}},
  \bibinfo{author}{\bibfnamefont{G.~Y.} \bibnamefont{Xiang}},
  \bibinfo{author}{\bibfnamefont{T.~C.} \bibnamefont{Ralph}}, \bibnamefont{and}
  \bibinfo{author}{\bibfnamefont{G.~J.} \bibnamefont{Pryde}},
  \bibinfo{journal}{Nat Phys} \textbf{\bibinfo{volume}{9}}, \bibinfo{pages}{23}
  (\bibinfo{year}{2013}), ISSN \bibinfo{issn}{1745-2473},
  \urlprefix\url{http://dx.doi.org/10.1038/nphys2469}.

\bibitem[{\citenamefont{Curty and Moroder}(2011)}]{CM11}
\bibinfo{author}{\bibfnamefont{M.}~\bibnamefont{Curty}} \bibnamefont{and}
  \bibinfo{author}{\bibfnamefont{T.}~\bibnamefont{Moroder}},
  \bibinfo{journal}{Phys. Rev. A} \textbf{\bibinfo{volume}{84}},
  \bibinfo{pages}{010304} (\bibinfo{year}{2011}),
  \urlprefix\url{http://link.aps.org/doi/10.1103/PhysRevA.84.010304}.

\bibitem[{\citenamefont{Clauser et~al.}(1969)\citenamefont{Clauser, Horne,
  Shimony, and Holt}}]{CHSH69}
\bibinfo{author}{\bibfnamefont{J.~F.} \bibnamefont{Clauser}},
  \bibinfo{author}{\bibfnamefont{M.~A.} \bibnamefont{Horne}},
  \bibinfo{author}{\bibfnamefont{A.}~\bibnamefont{Shimony}}, \bibnamefont{and}
  \bibinfo{author}{\bibfnamefont{R.~A.} \bibnamefont{Holt}},
  \bibinfo{journal}{Phys. Rev. Lett.} \textbf{\bibinfo{volume}{23}},
  \bibinfo{pages}{880} (\bibinfo{year}{1969}),
  \urlprefix\url{http://link.aps.org/doi/10.1103/PhysRevLett.23.880}.

\bibitem[{\citenamefont{Jin et~al.}(2015)\citenamefont{Jin, Takeoka, Takagi,
  Shimizu, and Sasaki}}]{JTTSS15}
\bibinfo{author}{\bibfnamefont{R.-B.} \bibnamefont{Jin}},
  \bibinfo{author}{\bibfnamefont{M.}~\bibnamefont{Takeoka}},
  \bibinfo{author}{\bibfnamefont{U.}~\bibnamefont{Takagi}},
  \bibinfo{author}{\bibfnamefont{R.}~\bibnamefont{Shimizu}}, \bibnamefont{and}
  \bibinfo{author}{\bibfnamefont{M.}~\bibnamefont{Sasaki}},
  \bibinfo{journal}{Scientific Reports} \textbf{\bibinfo{volume}{5}},
  \bibinfo{pages}{9333} (\bibinfo{year}{2015}),
  \urlprefix\url{http://dx.doi.org/10.1038/srep09333 10.1038/srep09333}.

\bibitem[{\citenamefont{Halder et~al.}(2007)\citenamefont{Halder, Beveratos,
  Gisin, Scarani, Simon, and Zbinden}}]{HBGSSZ07}
\bibinfo{author}{\bibfnamefont{M.}~\bibnamefont{Halder}},
  \bibinfo{author}{\bibfnamefont{A.}~\bibnamefont{Beveratos}},
  \bibinfo{author}{\bibfnamefont{N.}~\bibnamefont{Gisin}},
  \bibinfo{author}{\bibfnamefont{V.}~\bibnamefont{Scarani}},
  \bibinfo{author}{\bibfnamefont{C.}~\bibnamefont{Simon}}, \bibnamefont{and}
  \bibinfo{author}{\bibfnamefont{H.}~\bibnamefont{Zbinden}},
  \bibinfo{journal}{Nat Phys} \textbf{\bibinfo{volume}{3}},
  \bibinfo{pages}{692} (\bibinfo{year}{2007}), ISSN \bibinfo{issn}{1745-2473},
  \urlprefix\url{http://dx.doi.org/10.1038/nphys700}.

\bibitem[{\citenamefont{Pan et~al.}(1998)\citenamefont{Pan, Bouwmeester,
  Weinfurter, and Zeilinger}}]{PBWZ98}
\bibinfo{author}{\bibfnamefont{J.-W.} \bibnamefont{Pan}},
  \bibinfo{author}{\bibfnamefont{D.}~\bibnamefont{Bouwmeester}},
  \bibinfo{author}{\bibfnamefont{H.}~\bibnamefont{Weinfurter}},
  \bibnamefont{and}
  \bibinfo{author}{\bibfnamefont{A.}~\bibnamefont{Zeilinger}},
  \bibinfo{journal}{Phys. Rev. Lett.} \textbf{\bibinfo{volume}{80}},
  \bibinfo{pages}{3891} (\bibinfo{year}{1998}),
  \urlprefix\url{http://link.aps.org/doi/10.1103/PhysRevLett.80.3891}.

\bibitem[{\citenamefont{Lim et~al.}(2013)\citenamefont{Lim, Portmann,
  Tomamichel, Renner, and Gisin}}]{LPTRG13}
\bibinfo{author}{\bibfnamefont{C.~C.~W.} \bibnamefont{Lim}},
  \bibinfo{author}{\bibfnamefont{C.}~\bibnamefont{Portmann}},
  \bibinfo{author}{\bibfnamefont{M.}~\bibnamefont{Tomamichel}},
  \bibinfo{author}{\bibfnamefont{R.}~\bibnamefont{Renner}}, \bibnamefont{and}
  \bibinfo{author}{\bibfnamefont{N.}~\bibnamefont{Gisin}},
  \bibinfo{journal}{Phys. Rev. X} \textbf{\bibinfo{volume}{3}},
  \bibinfo{pages}{031006} (\bibinfo{year}{2013}),
  \urlprefix\url{http://link.aps.org/doi/10.1103/PhysRevX.3.031006}.

\bibitem[{\citenamefont{Acin et~al.}(2006)\citenamefont{Acin, Massar, and
  Pironio}}]{AMP06}
\bibinfo{author}{\bibfnamefont{A.}~\bibnamefont{Acin}},
  \bibinfo{author}{\bibfnamefont{S.}~\bibnamefont{Massar}}, \bibnamefont{and}
  \bibinfo{author}{\bibfnamefont{S.}~\bibnamefont{Pironio}},
  \bibinfo{journal}{New Journal of Physics} \textbf{\bibinfo{volume}{8}},
  \bibinfo{pages}{126} (\bibinfo{year}{2006}),
  \urlprefix\url{http://stacks.iop.org/1367-2630/8/i=8/a=126}.

\bibitem[{\citenamefont{Cirel'son}(1980)}]{Cirel80}
\bibinfo{author}{\bibfnamefont{B.}~\bibnamefont{Cirel'son}},
  \bibinfo{journal}{Letters in Mathematical Physics}
  \textbf{\bibinfo{volume}{4}}, \bibinfo{pages}{93} (\bibinfo{year}{1980}),
  ISSN \bibinfo{issn}{0377-9017},
  \urlprefix\url{http://dx.doi.org/10.1007/BF00417500}.

\bibitem[{\citenamefont{Devetak and Winter}(2005)}]{DW05}
\bibinfo{author}{\bibfnamefont{I.}~\bibnamefont{Devetak}} \bibnamefont{and}
  \bibinfo{author}{\bibfnamefont{A.}~\bibnamefont{Winter}},
  \bibinfo{journal}{Proceedings of the Royal Society of London A: Mathematical,
  Physical and Engineering Sciences} \textbf{\bibinfo{volume}{461}},
  \bibinfo{pages}{207} (\bibinfo{year}{2005}), ISSN \bibinfo{issn}{1364-5021}.

\bibitem[{\citenamefont{Caprara~Vivoli
  et~al.}(2015)\citenamefont{Caprara~Vivoli, Sekatski, Bancal, Lim,
  Christensen, Martin, Thew, Zbinden, Gisin, and Sangouard}}]{VSBLC15}
\bibinfo{author}{\bibfnamefont{V.}~\bibnamefont{Caprara~Vivoli}},
  \bibinfo{author}{\bibfnamefont{P.}~\bibnamefont{Sekatski}},
  \bibinfo{author}{\bibfnamefont{J.-D.} \bibnamefont{Bancal}},
  \bibinfo{author}{\bibfnamefont{C.~C.~W.} \bibnamefont{Lim}},
  \bibinfo{author}{\bibfnamefont{B.~G.} \bibnamefont{Christensen}},
  \bibinfo{author}{\bibfnamefont{A.}~\bibnamefont{Martin}},
  \bibinfo{author}{\bibfnamefont{R.~T.} \bibnamefont{Thew}},
  \bibinfo{author}{\bibfnamefont{H.}~\bibnamefont{Zbinden}},
  \bibinfo{author}{\bibfnamefont{N.}~\bibnamefont{Gisin}}, \bibnamefont{and}
  \bibinfo{author}{\bibfnamefont{N.}~\bibnamefont{Sangouard}},
  \bibinfo{journal}{Phys. Rev. A} \textbf{\bibinfo{volume}{91}},
  \bibinfo{pages}{012107} (\bibinfo{year}{2015}),
  \urlprefix\url{http://link.aps.org/doi/10.1103/PhysRevA.91.012107}.

\bibitem[{\citenamefont{Jin et~al.}(2014)\citenamefont{Jin, Shimizu, Wakui,
  Fujiwara, Yamashita, Miki, Terai, Wang, and Sasaki}}]{JSWF14}
\bibinfo{author}{\bibfnamefont{R.-B.} \bibnamefont{Jin}},
  \bibinfo{author}{\bibfnamefont{R.}~\bibnamefont{Shimizu}},
  \bibinfo{author}{\bibfnamefont{K.}~\bibnamefont{Wakui}},
  \bibinfo{author}{\bibfnamefont{M.}~\bibnamefont{Fujiwara}},
  \bibinfo{author}{\bibfnamefont{T.}~\bibnamefont{Yamashita}},
  \bibinfo{author}{\bibfnamefont{S.}~\bibnamefont{Miki}},
  \bibinfo{author}{\bibfnamefont{H.}~\bibnamefont{Terai}},
  \bibinfo{author}{\bibfnamefont{Z.}~\bibnamefont{Wang}}, \bibnamefont{and}
  \bibinfo{author}{\bibfnamefont{M.}~\bibnamefont{Sasaki}},
  \bibinfo{journal}{Opt. Express} \textbf{\bibinfo{volume}{22}},
  \bibinfo{pages}{11498} (\bibinfo{year}{2014}),
  \urlprefix\url{http://www.opticsexpress.org/abstract.cfm?URI=oe-22-10-11498}.

\bibitem[{\citenamefont{Grice and Walmsley}(1997)}]{GW97}
\bibinfo{author}{\bibfnamefont{W.~P.} \bibnamefont{Grice}} \bibnamefont{and}
  \bibinfo{author}{\bibfnamefont{I.~A.} \bibnamefont{Walmsley}},
  \bibinfo{journal}{Physical Review A} \textbf{\bibinfo{volume}{56}},
  \bibinfo{pages}{1627} (\bibinfo{year}{1997}), ISSN \bibinfo{issn}{1050-2947}.

\bibitem[{\citenamefont{Law et~al.}(2000)\citenamefont{Law, Walmsley, and
  Eberly}}]{LWE00}
\bibinfo{author}{\bibfnamefont{C.~K.} \bibnamefont{Law}},
  \bibinfo{author}{\bibfnamefont{I.~A.} \bibnamefont{Walmsley}},
  \bibnamefont{and} \bibinfo{author}{\bibfnamefont{J.~H.}
  \bibnamefont{Eberly}}, \bibinfo{journal}{Physical Review Letters}
  \textbf{\bibinfo{volume}{84}}, \bibinfo{pages}{5304} (\bibinfo{year}{2000}),
  ISSN \bibinfo{issn}{0031-9007}.

\bibitem[{\citenamefont{Mosley et~al.}(2008)\citenamefont{Mosley, Lundeen,
  Smith, Wasylczyk, U'Ren, Silberhorn, and Walmsley}}]{MLSWUSW08}
\bibinfo{author}{\bibfnamefont{P.~J.} \bibnamefont{Mosley}},
  \bibinfo{author}{\bibfnamefont{J.~S.} \bibnamefont{Lundeen}},
  \bibinfo{author}{\bibfnamefont{B.~J.} \bibnamefont{Smith}},
  \bibinfo{author}{\bibfnamefont{P.}~\bibnamefont{Wasylczyk}},
  \bibinfo{author}{\bibfnamefont{A.~B.} \bibnamefont{U'Ren}},
  \bibinfo{author}{\bibfnamefont{C.}~\bibnamefont{Silberhorn}},
  \bibnamefont{and} \bibinfo{author}{\bibfnamefont{I.~a.}
  \bibnamefont{Walmsley}}, \bibinfo{journal}{Physical Review Letters}
  \textbf{\bibinfo{volume}{100}}, \bibinfo{pages}{1} (\bibinfo{year}{2008}),
  ISSN \bibinfo{issn}{00319007}, \eprint{0711.1054}.

\bibitem[{\citenamefont{Takeoka et~al.}(2015)\citenamefont{Takeoka, Jin, and
  Sasaki}}]{TJS15}
\bibinfo{author}{\bibfnamefont{M.}~\bibnamefont{Takeoka}},
  \bibinfo{author}{\bibfnamefont{R.-b.} \bibnamefont{Jin}}, \bibnamefont{and}
  \bibinfo{author}{\bibfnamefont{M.}~\bibnamefont{Sasaki}},
  \bibinfo{journal}{New Journal of Physics} \textbf{\bibinfo{volume}{17}},
  \bibinfo{pages}{43030} (\bibinfo{year}{2015}), ISSN
  \bibinfo{issn}{1367-2630},
  \urlprefix\url{http://dx.doi.org/10.1088/1367-2630/17/4/043030}.

\bibitem[{\citenamefont{Adesso et~al.}(2014)\citenamefont{Adesso, Ragy, and
  Lee}}]{ARL14}
\bibinfo{author}{\bibfnamefont{G.}~\bibnamefont{Adesso}},
  \bibinfo{author}{\bibfnamefont{S.}~\bibnamefont{Ragy}}, \bibnamefont{and}
  \bibinfo{author}{\bibfnamefont{A.~R.} \bibnamefont{Lee}},
  \bibinfo{journal}{Open Systems \& amp; Information Dynamics}
  \textbf{\bibinfo{volume}{21}}, \bibinfo{pages}{1440001}
  (\bibinfo{year}{2014}).

\bibitem[{\citenamefont{Weedbrook et~al.}(2012)\citenamefont{Weedbrook,
  Pirandola, Garcia-Patron, Cerf, Ralph, Shapiro, and Lloyd}}]{WPGCRSL12}
\bibinfo{author}{\bibfnamefont{C.}~\bibnamefont{Weedbrook}},
  \bibinfo{author}{\bibfnamefont{S.}~\bibnamefont{Pirandola}},
  \bibinfo{author}{\bibfnamefont{R.}~\bibnamefont{Garcia-Patron}},
  \bibinfo{author}{\bibfnamefont{N.~J.} \bibnamefont{Cerf}},
  \bibinfo{author}{\bibfnamefont{T.~C.} \bibnamefont{Ralph}},
  \bibinfo{author}{\bibfnamefont{J.~H.} \bibnamefont{Shapiro}},
  \bibnamefont{and} \bibinfo{author}{\bibfnamefont{S.}~\bibnamefont{Lloyd}},
  \bibinfo{journal}{Reviews of Modern Physics} \textbf{\bibinfo{volume}{84}},
  \bibinfo{pages}{621} (\bibinfo{year}{2012}), \bibinfo{note}{arXiv:1110.3234}.

\end{thebibliography}
\end{acknowledgments}

\section*{Appendix: The Characteristic Function Approach to Photonic Quantum
information processing\label{sec:Characteristic-Function-Approach}}

In continuous-variable quantum information processing, there exist
powerful tools based on the characteristic functions of quantum states,
which are particularly useful when dealing with Gaussian states and
Gaussian operations \cite{ARL14,WPGCRSL12}. Since entangled photon
pairs, in practice, are post-selected from continuous-variable sources
such as SPDCs, these tools lend themselves rather naturally for an
easy and exact treatment of photonic quantum information processing
tasks (cf. \cite{TJS15}) without the need for any approximations.
Here we present a brief review of these tools for the convenience
of the reader. (For a more comprehensive review, see \cite{ARL14,WPGCRSL12}.)

Consider $N$ Bosonic modes associated with a tensor product Hilbert
space $\mathcal{{H}}^{\otimes N}=\otimes_{j=1}^{N}\mathcal{H}_{j}$
, where each $\mathcal{H}_{j}$ is an infinite-dimensional Hilbert
space. Corresponding to each mode is a pair of field operators $\hat{{a}}_{j}$
and $\hat{{a}}_{j}^{\dagger}$\---the annihilation and creation operators\---which
satisfy the canonical commutation relation given by
\begin{equation}
\left[\hat{{a}}_{j},\hat{{a}}_{k}^{\dagger}\right]=\delta_{jk}.
\end{equation}
It is common to define the quadrature operators of a bosonic mode
as
\begin{align}
\hat{{x}}_{j} & =\frac{{1}}{\sqrt{{2}}}\left(\hat{{a}}_{j}+\hat{{a}}_{j}^{\dagger}\right),\\
\hat{{p}}_{j} & =\frac{{1}}{\sqrt{{2}}i}\left(\hat{{a}}_{j}-\hat{{a}}_{j}^{\dagger}\right),
\end{align}
where these operators can be verified to obey the commutation relation
\begin{equation}
\left[\hat{{x}}_{j},\hat{{p}}_{k}\right]=i\delta_{jk}.
\end{equation}
(Note that we choose as a convention $\hbar=1$.)

Let $\hat{{\rho}}$ be a density operator defined on $\mathcal{{H}}^{\otimes N}$,
which represents a quantum state in the $N$-mode Hilbert space. The
characteristic function of $\hat{\rho}$ is defined to be
\begin{equation}
\chi\left(\xi\right)=\operatorname{Tr}\left\{ \hat{{\rho}}\hat{\mathcal{W}}\left(\xi\right)\right\} ,
\end{equation}
where
\begin{equation}
\hat{\mathcal{W}}\left(\xi\right)=\exp\left(-i\xi^{T}\hat{{R}}\right)
\end{equation}
is known as the Weyl operator, and
\begin{align}
\hat{R} & =\left[\hat{x}_{1},\ldots\hat{{x}}_{N},\hat{p}_{1},\ldots\hat{{p}}_{N}\right],\\
\xi & =\left[\xi_{1},\ldots,\xi_{2N}\right],\:\xi_{i}\in\mathbb{R}\forall i.
\end{align}

\subsection{Gaussian States}

A Gaussian state is a quantum state whose characteristic function
is Gaussian, i.e., of the form:
\begin{equation}
\chi\left(x\right)=\exp\left[-\frac{{1}}{4}x^{\operatorname{T}}\gamma x-i\operatorname{d}^{\operatorname{T}}x\right],
\end{equation}
where $\gamma$ is a $2n\times2n$ matrix called the covariance matrix
and $\operatorname{d}$ is a $2n$-dimensional vector known as the
displacement vector. The simplest example of a Gaussian state is the
coherent state
\begin{align}
\left|\alpha\right\rangle  & =\exp\left(\alpha a^{\dagger}-\alpha^{*}a\right)\left|0\right\rangle ,\;\alpha=\left|\alpha\right|\exp\left(i\phi\right),\\
 & =\exp\left(-\frac{\left|\alpha\right|^{2}}{2}\right)\sum_{n=0}^{\infty}\frac{\alpha^{n}}{\sqrt{{n}!}}\left|n\right\rangle .
\end{align}
Its covariance matrix is the $2\times2$ identity matrix, and displacement
vector is
\begin{equation}
\alpha=\sqrt{{2}}\begin{pmatrix}\begin{array}[t]{c}
\operatorname{Re}\left(\alpha\right)\\
\operatorname{Im}\left(\alpha\right)
\end{array}\end{pmatrix}.
\end{equation}
Another common example of a Gaussian state is the two-mode squeezed
vacuum state, which is generated by an SPDC source
\begin{align}
\left|\xi\right\rangle  & =\exp\left(\xi a^{\dagger}b^{\dagger}-\xi^{*}ab\right)\left|0\right\rangle _{a}\otimes\left|0\right\rangle _{b}\\
 & =\frac{{1}}{\cosh r}\sum_{n=0}^{\infty}\left(\exp\left({i\theta}\right)\tanh r\right)^{n}\left|n\right\rangle _{a}\otimes\left|n\right\rangle _{b},
\end{align}
where $\xi=r\exp\left(i\theta\right)$ is the squeezing parameter.
The TMSV has zero displacement and a covariance matrix given by
\begin{equation}
\gamma^{\operatorname{TMSV}}\left(\mu\right)=\begin{pmatrix}\gamma^{+}\left(\mu\right) & 0\\
0 & \gamma^{-}\left(\mu\right)
\end{pmatrix},
\end{equation}
where $\mu$ is the average photon number in each mode of the state
and
\begin{equation}
\gamma^{\pm}\left(\mu\right)=\begin{pmatrix}2\mu+1 & \pm2\sqrt{{\mu\left(\mu+1\right)}}\\
\pm2\sqrt{{\mu\left(\mu+1\right)}} & 2\mu+1
\end{pmatrix}.
\end{equation}

A very convenient property of the characteristic function representation
of a multimode Gaussian state is that the reduced state on any subsystem
is simply given by the corresponding sub-matrix of the displacement
vector and the covariance matrix of the full state. For example, consider
the TMSV. The reduced state on any one of the two modes is a thermal
state
\begin{equation}
\rho^{\operatorname{th}}=\sum_{n=0}^{\infty}\frac{{\mu^{n}}}{\left(\mu+1\right)^{n+1}}\left|n\right\rangle \left\langle n\right|,
\end{equation}
whose covariance matrix is given by
\begin{equation}
\gamma^{\operatorname{th}}=\begin{pmatrix}2\mu+1 & 0\\
0 & 2\mu+1
\end{pmatrix},
\end{equation}
which is precisely what the corresponding sub-matrix of $\gamma^{\operatorname{TMSV}}$
is.

\subsection{Gaussian operations}

By a quantum operation, we mean a linear map $\mathcal{E}:\rho\rightarrow\mathcal{E}\left(\rho\right)$
(where $\rho$ is a quantum state, i.e., $\rho\geq0$ and $\operatorname{Tr}\left(\rho\right)=1$)),
which is completely positive, i.e. $\left(\operatorname{id}\otimes\mathcal{E}\right)\left(\sigma\otimes\rho\right)$
is also a valid quantum state for all positive operators $\sigma$,
and trace reducing, i.e., $0\leq\operatorname{Tr}\left(\mathcal{E}\left(\rho\right)\right)\leq1$.
A quantum operation is called a quantum channel if it is trace preservation,
i.e., $\operatorname{Tr}\left(\mathcal{E}\left(\rho\right)\right)=1$.
Further, the special case of quantum channels that are reversible
are the unitary transformations $U^{-1}=U^{\dagger}$, which transform
a quantum state $\rho$ as $\rho\rightarrow U\rho U^{\dagger}$.

A quantum operation is called a Gaussian operation if it maps Gaussian
states to Gaussian states. Also likewise, Gaussian unitaries are defined
to be unitaries that map Gaussian states to Gaussian states. The action
of a Gaussian unitary $U$ on a state $\rho$ can be easily described
easily by a corresponding real symplectic transformation $S$ on the
covariance matrix $\gamma$ and the displacement vector $\operatorname{d}$
of the state
\begin{equation}
\gamma\rightarrow S^{\operatorname{T}}\gamma S,\;\operatorname{d}\rightarrow S^{\operatorname{T}}\operatorname{d}.
\end{equation}

The symplectic transformation corresponding to a simple phase shift
unitary on a single mode is given by
\begin{equation}
R\left(\phi\right)=\begin{pmatrix}\cos\phi & \sin\phi\\
-\sin\phi & \cos\phi
\end{pmatrix}.
\end{equation}
Likewise, that corresponding to a beamsplitter of transmittivity $t$
between two modes is given by
\begin{equation}
S\left(t\right)=\begin{pmatrix}\sqrt{{t}} & \sqrt{{1-t}} & 0 & 0\\
-\sqrt{{1-t}} & \sqrt{{t}} & 0 & 0\\
0 & 0 & \sqrt{{t}} & \sqrt{{1-t}}\\
0 & 0 & -\sqrt{{1-t}} & \sqrt{{t}}
\end{pmatrix}.
\end{equation}

\subsection{Photodetectors\label{sub:Photodetectors}}

We consider on-off photodetectors, meaning that the detectors simply
distinguish between vacuum and not vacuum. These detectors can be
represented as the following positive operator valued measure (POVM):
\begin{align}
\Pi_{0} & =\left|0\right\rangle \left\langle 0\right|,\nonumber \\
\Pi_{1} & =\sum_{n=1}^{\infty}\left|n\right\rangle \left\langle n\right|=I-\Pi_{0}.\label{eq:on-off-det}
\end{align}

When a single-mode Gaussian state $\rho$ with characteristic function
$\chi_{\rho}\left(x\right)=\exp\left(-\frac{{1}}{4}x^{\operatorname{T}}\gamma x\right)$
is measured using a on-off photodetector, the probability of detecting
photons (``on'' outcome) is given by
\begin{align}
p_{1} & =\operatorname{Tr}\left(\rho\Pi_{1}\right)=1-\operatorname{Tr}\left(\rho\Pi_{0}\right)\nonumber \\
 & =1-\frac{{1}}{2\pi}\int dx\,\chi_{\rho}\left(x\right)\chi_{\left|0\right\rangle \left\langle 0\right|}\left(-x\right)\nonumber \\
 & =1-\frac{{1}}{2\pi}\int dx\,\exp\left(-\frac{{1}}{4}x^{\operatorname{T}}\left(\gamma+I\right)x\right)\nonumber \\
 & =1-\frac{{2}}{\sqrt{{\operatorname{det}\left(\gamma+I\right)}}}.
\end{align}
Likewise, when an m-mode Gaussian state is measured using on-off
photodetectors in all the modes, the probability of coincidence detection
(``on'' outcome in all the modes) is given by
\begin{equation}
p_{\operatorname{coinc.}}=\sum_{\tau\in\mathcal{{P\left(K\right)}}}\left(-1\right)^{\left|\tau\right|}\frac{{2^{\left|\tau\right|}}}{\sqrt{{\operatorname{det}\left(\gamma^{\left(\tau\right)}+I_{\left|\tau\right|}\right)}}},
\end{equation}
where $\mathcal{{K}}$ is a set consisting of the m modes, $\mathcal{{P\left(K\right)}}$
is the powerset of $\mathcal{{K}}$\---meaning the set of all subsets
of $\mathcal{{K}}$, $\gamma^{\left(\tau\right)}$, e.g., is the covariance
matrix of the reduced state on the modes in element $\tau$, and $I_{\left|\tau\right|}$
is the identity matrix of dimension $\left|\tau\right|$.

\subsection{Imperfections in the channel and the detectors}

The primary imperfection in the optical channel is photon loss, e.g.,
losses in transmission and in coupling between media. It is known
that the lossy optical channel is a Gaussian channel. And a typical
model for the channel is a pure loss bosonic channel, which is a beamsplitter
transformation of transmittivity $t$ between the lossy mode and a
vacuum mode. The action of the lossy optical channel on the state
of a mode with covariance matrix $\gamma$ can be described as
\begin{equation}
\mathcal{\mathcal{L}}^{t}:\gamma\rightarrow K^{\operatorname{T}}\gamma K+\alpha,
\end{equation}
where $K=\sqrt{{t}}I$ and $\alpha=\left(1-t\right)I$.

The imperfections in the on-off photodetectors include (a) photon
loss\---this is modeled as a lossy channel of the above type followed
by a lossless detector, (b) dark counts\---these are modeled by amending
the detector POVM elements as
\begin{align}
\Pi_{0}\left(\nu\right) & =\left(1-\nu\right)\left|0\right\rangle \left\langle 0\right|,\\
\Pi_{1}\left(\nu\right) & =I-\Pi_{0}\left(\nu\right),
\end{align}
where $\nu$ is the dark count probability.
\end{document}